\documentclass[acmsmall,screen]{acmart}
\setcopyright{none}
\copyrightyear{2018}
\acmYear{2018}
\acmDOI{XXXXXXX.XXXXXXX}

\acmJournal{JACM}
\acmVolume{37}
\acmNumber{4}
\acmArticle{111}
\acmMonth{8}

\newcommand{\fix}[1]{\textcolor{red}{\textbf{\emph{#1}}}}

\usepackage{booktabs}   
\usepackage[labelformat=simple]{subcaption} 

\usepackage{comment}
\usepackage{multirow}
\usepackage[ruled,vlined,linesnumbered]{algorithm2e}
\usepackage{pifont}
\usepackage{makecell}
\newcommand{\cmark}{\ding{51}}%
\newcommand{\xmark}{\ding{55}}%

\usepackage{tikz}
\usetikzlibrary{intersections,calc,patterns}

\usepackage{pgfplotstable}

\usepackage{pgfplots}

\pgfplotsset{compat=1.17} 

\usepackage{listings}
\lstdefinelanguage{z3}{
    alsoletter=-,morekeywords={assert, check-sat,and,or,not},morecomment=[l]{;}
}

\begin{document}

\title{Polyhedral Specification and Code Generation of Sparse Tensor Contraction with Co-Iteration}



\author{Tuowen Zhao}
\email{ztuowen@gmail.com}
\affiliation{%
  \institution{University of Utah}
  \city{Salt Lake City}
  \state{Utah}
  \country{USA}
}

\author{Tobi Popoola}
\email{tobipopoola@u.boisestate.edu}
\affiliation{%
  \institution{Boise State University}
  \city{Boise}
  \state{Idaho}
  \country{USA}
}

\author{Mary Hall}
\email{mhall@cs.utah.edu}
\affiliation{%
  \institution{University of Utah}
  \city{Salt Lake City}
  \state{Utah}
  \country{USA}
}

\author{Catherine Olschanowsky}
\email{catherineolschan@boisestate.edu}
\affiliation{
\institution{Boise State University}
\city{Boise}
\state{Idaho}
\country{USA}
}

\author{Michelle Mills Strout}
\email{mstrout@cs.arizona.edu}
\affiliation{
\institution{University of Arizona}
\city{Tucson}
\state{Arizona}
\country{USA}
}        


\begin{abstract}
This paper presents a code generator for sparse tensor contraction computations. 
It leverages a mathematical representation of loop nest computations in the sparse polyhedral framework (SPF), which extends the polyhedral model to support non-affine computations, such as arise in sparse tensors. 
SPF is extended to perform layout specification, optimization, and code generation of sparse tensor code: 1) we develop a polyhedral layout specification that decouples iteration spaces for layout and computation; and, 2) we develop efficient co-iteration of sparse tensors by combining polyhedra scanning over the layout of one sparse tensor with the synthesis of code to \emph{find} corresponding elements in other tensors through an SMT solver.  

We compare the generated code with that produced by a state-of-the-art tensor compiler, TACO. 
We achieve on average 1.63$\times$ faster parallel performance 
than TACO on sparse-sparse co-iteration and describe how to improve that to 2.72$\times$ average speedup by switching the find algorithms.
We also demonstrate that decoupling iteration spaces of layout and computation 
enables additional 
layout and computation combinations to be supported.
\end{abstract}



\maketitle

\section{Introduction}

Tensor contractions are found in a wide variety of computations in data science, machine learning, and finite element methods~\citep{Bader2008,tvm,tensorflow,einstein,NWChem}. Sparse tensors are tensors that contain a large number of zero values that have been compressed out to save memory and avoid unnecessary computation. 
A \emph{layout} is used to represent a sparse tensor, which includes the nonzero values and a set of  auxiliary data structures that relate nonzero values to their indices in the computation. Many sparse tensor layouts have been introduced to improve performance under different algorithmic contexts, sparsity patterns, and for different target architectures (for examples, see survey by~\citet{langr2016evaluation}).

A sparse tensor layout can be thought of as a \emph{physical} description of the sparse tensor -- how it is ordered in memory and the requisite auxiliary data structures that define its meaning.  The \emph{logical} view of the sparse tensor is its dense form, which is usually prohibitively large to represent in memory; the logical abstraction of the nonzeros must be preserved by the physical layout.  

To optimize both computation and layout of sparse tensors, several sparse tensor compilers have been developed that generate optimized code from a dense description of the computation, using a sparse physical layout of the tensor~\cite{aartthesis,kotlyar1997relational,taco}.  
Most recently, the Tensor Algebra Compiler (TACO)~\cite{taco} uses \emph{level formats}~\cite{tacoformat}
to describe the physical storage of different index dimensions of a tensor, with each level also associated with an index dimension in the tensor computation.
This layout description and the formulation using \emph{merge lattices} allows dimensions from multiple sparse tensors to be \emph{co-iterated}, which refers to matching coordinates of nonzeros in one sparse tensor to those in another sparse tensor. For example, in sparse dot product, if coordinate $p$ is nonzero for one of the vectors, the element-wise product is nonzero if and only if $p$ is also nonzero in the other vector.  TACO's support of co-iteration extends the applicability of tensor compilers.

We observe that the use of level formats and merge lattices couples the 
logical (computation's coordinate space) and physical (layout's position space) 
dimensions and their associated  
iteration ranges; consequently, this approach requires that each level in the layout must refer to a distinct index in the computation.  Level formats are unable to directly support blocked layouts 
such as block compressed sparse row (BCSR), which have additional physical dimensions not present in the computation.  
Additionally, generalizations of contraction that use the same loop index for multiple levels, e.g., computations along a matrix diagonal, cannot be directly supported due to conflicts in iteration ranges.  

In this paper, we separate the physical layout of the sparse tensor (layout's position space) from logical indices (computation's coordinate space), by preserving indices from both spaces and describing a mathematical relation between them.
For this purpose, our representation extends the 
polyhedral model~\citep{feautrier1992some1,feautrier1992some2,griebl98,Shen98,quillere,girbal06,pluto,tobias2015}, an abstraction used to 
represent integer sets and compose optimizations on loop nest 
computations, and compose computation with storage mappings~\citep{Lefebvre:1998:ASM,StroutEtAl98,Polyhedral2000,Thies2007}.  A rich set of affine code transformations can be described using the polyhedral framework and related mathematical representations including locality optimization~\citep{supernode,wolfe89,wolflam91,ramanujam92}, automatic and semi-automatic parallelization~\citep{supernode,girbal06,pluto,chill,cudachill}, and auto-distribution~\citep{kelly_optimization_1998,tiramisu}. To support sparse computation involving
non-affine loop bounds and indirect accesses in subscript expressions (e.g., A[B[i]]), our approach employs techniques from the Sparse Polyhedral Framework (SPF)~\citep{Strout16,Strout18}, which uses \emph{uninterpreted functions} to represent values of 
auxiliary \emph{index arrays} that are only known at runtime. 

Prior work using SPF has not presented a solution to co-iteration of multiple sparse tensors~\cite{StroutPLDI03,anandpldi,mahdiarrayprop,Strout18}.  
In this paper, we extend SPF to support co-iteration by generating code that iterates over the layout of one sparse tensor's nonzeros and looks up corresponding nonzeros in other tensors with \emph{find} operations.     

This paper makes the following contributions: 
1) We describe a sparse tensor layout as a relation between the logical and physical space, which supports layouts that cannot be described when coordinate and position spaces are coupled;
2) We extend SPF to generate efficient co-iteration code 
    through a combination of polyhedra scanning and code synthesis using a satisfiability modulo theories (SMT) solver.
3) We show how the use of SPF facilitates parallelization and composition of transformations, demonstrated by deriving data dependence relations and tiling the code;
4) We compare the proposed method with a state-of-the-art tensor algebra compiler, TACO. On sparse-sparse co-iteration in the sparse-matrix times sparse-vector experiments, we achieve on average 1.63$\times$ speedup against TACO using a find algorithm comparable to TACO's co-iteration implementation and improve that to 2.72$\times$ average speedup when we switch between find algorithms. We can even express cases when the input and output share sparsity structure such as in the mode-1 tensor-times-matrix computation, where we are able to achieve 4.27$\times$ average speedup on real-world 3D tensors.

\section{Background and Overview}
This section formulates tensor contraction code generation in the polyhedral framework for dense tensors and demonstrates how the sparse polyhedral framework represents the index arrays arising from sparse tensors.  The end of the section gives an overview for the remainder of the paper.

\subsection{Tensor Contraction}

Tensor contractions can be expressed using 
the \emph{tensor index notation} described by~\citet{tensorindex}, where dimensionality of input tensors is contracted along one or more dimensions.  Examples from linear algebra include dot product, 
$y = A(i)*B(i)$, matrix-vector multiplication, $y(i) = A(i,j) * B(j)$, and 
matrix-matrix multiplication,  $y(i,j) = A(i,k) * B(k,j)$. 
Higher-dimensional tensor contractions are common occurrences in machine learning and the finite element method.
This notation expresses the accesses of the input and output tensors in the computation.  
Indices only appearing on the right hand side, such as $k$ in matrix-matrix multiplication, are commonly referred to as \emph{summation} or \emph{contraction} indices and introduce a data dependence; indices appearing on both sides, such as $i$ and $j$, are commonly called the \emph{external} or \emph{free} indices. Because the contraction index $k$
iterates over the second dimension of $A$ and the first dimension of $B$ at the same time, this behavior is 
referred to as co-iteration.  


\subsection{Polyhedral Framework}

\begin{figure*}[t]
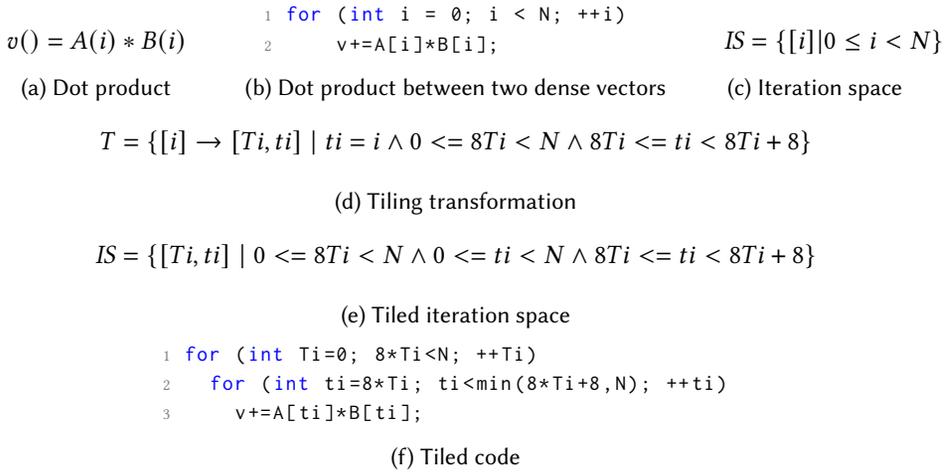

\centering

\begin{subfigure}[b]{0.2\textwidth}
\centering
\begin{tabular}{c}
$v() = A(i) * B(i)$\\
\end{tabular}
\caption{Dot product}
\label{fig:dot-product}
\end{subfigure}
~
\begin{subfigure}[b]{0.475\textwidth}
\centering
\begin{tabular}{c}
\begin{lstlisting}[language=C++,escapechar=|]
for (int i = 0; i < N; ++i)
    v+=A[i]*B[i];|\label{line:d-is}|
\end{lstlisting}
\end{tabular}
\caption{Dot product between two dense vectors}
\label{fig:dense-dot}
\end{subfigure}
~
\begin{subfigure}[b]{0.2\textwidth}
\centering
\begin{tabular}{c}
    $\mathit{IS} = \{ [i] | 0 \leq i < N \}$\\
\end{tabular}
\caption{Iteration space}
\label{fig:dense-is}
\end{subfigure}

\begin{subfigure}[b]{0.8\textwidth}
\centering
\begin{align*}
    T = \{ [i] \rightarrow [Ti,ti] \;|\; ti = i \wedge 0 <= 8Ti < N \wedge 8Ti <= ti < 8Ti + 8  \}
\end{align*}
\caption{Tiling transformation}
\label{fig:tiling-trans}
\end{subfigure}

\begin{subfigure}[b]{0.8\textwidth}
\centering
\begin{align*}
    \mathit{IS} = \{ [Ti,ti] \;|\; 0 <= 8Ti < N \wedge 0 <= ti < N \wedge 8Ti <= ti < 8Ti + 8  \}
\end{align*}
\caption{Tiled iteration space}
\label{fig:tiling-is}
\end{subfigure}

\begin{subfigure}[b]{0.6\textwidth}
\centering
\begin{tabular}{c}
\begin{lstlisting}[language=C++,escapechar=|]
for (int Ti=0; 8*Ti<N; ++Ti)
  for (int ti=8*Ti; ti<min(8*Ti+8,N); ++ti)
    v+=A[ti]*B[ti];
\end{lstlisting}\\[10pt]
\end{tabular}
\caption{Tiled code}
\label{fig:tiling-code}
\end{subfigure}

\caption{Dot product between dense vectors.}

\label{fig:dense-dot-all}
\end{figure*}

Polyhedral frameworks describe 
the instances of a statement's execution in a loop nest
as a set of lattice points of polyhedra.  Polyhedral compilers were designed to support computations that are in the \emph{affine domain}, where loop bounds and subscript expressions are integer linear functions of loop indices and constants.
Polyhedra are specified by a \emph{Presburger formula} on index variables through affine constraints, logical operators, and existential operators. When specified this way, this set of lattice points are also called a \emph{Presburger set}. Presburger sets and relations (Definition~\ref{def:relation}) are denoted using capital letters such as $A, R, P, T, Q$ and for iteration space, $\mathit{IS}$. Presburger set $R_{x_1, x_2, \dots, x_d}$, with set variables $(x_1, x_2, \dots, x_d)$ and Presburger formula, $P$, is written as follows.

\begin{equation*}
    R_{x_1, x_2, \dots, x_d} = \{[x_1, \dots, x_d] | P \}
\end{equation*}

Consider the dot product over two dense tensors in
Fig.~\ref{fig:dense-dot-all}, expressed in tensor notation in Fig.~\ref{fig:dot-product} with corresponding C code in Fig.~\ref{fig:dense-dot}. We describe 
the \emph{iteration space} for the statement on line~\ref{line:d-is} by the polyhedron of Fig.~\ref{fig:dense-is}\footnote{Auxiliary indices may be introduced to differentiate different statements in the same loop level for imperfectly nested loop nests.}.
A statement macro is used to represent the statement at line 2 as a function of loop index $i$.

An important capability of polyhedral frameworks is the ability to represent transformations on loop nests as affine mappings on the iteration spaces.  
For example, tiling the $i$ loop iteration by 8 can be represented by Fig.~\ref{fig:tiling-trans}. When $T$ is applied to the iteration space of Fig.~\ref{fig:dense-is}, the resulting 
iteration space is Fig.~\ref{fig:tiling-is}. A sequence of transformations are applied by composing the mappings. 
Polyhedral compilers generate code by performing polyhedral scanning~\citep{scanning,quillere,Chenpldi2012}. Scanning produces constraints on each loop index from 
the iteration space description. 
These constraints are directly translated to for loops and if conditions in the generated code. 
Loop indices in the transformed statement of Line~\ref{line:d-is} are substituted using the inverse mapping resulting in code shown in Fig.~\ref{fig:tiling-code}.
 
Some of the common operations on Presburger sets and relations are used in this paper.

\begin{definition}
Intersection between Presburger sets, $R = R_1 \cap R_2$:
$$\mathbf{s} \in R \iff \mathbf{s} \in R_1 \wedge \mathbf{s} \in R_2$$
\end{definition}

\begin{definition}
A Presburger relation denotes a binary relation between the input set of indices, $\mathbf{i}$, and output set of indices, $\mathbf{o}$, described as $R_{\mathbf{i} \rightarrow \mathbf{o}} = \{ \mathbf{i} \rightarrow \mathbf{o} | P_{\mathbf{i} \rightarrow \mathbf{o}}\}$.
\label{def:relation}
\end{definition}

\begin{definition}
Compositions are between two Presburger relations, $R_{\mathbf{x} \rightarrow \mathbf{o}} = R_{\mathbf{i} \rightarrow \mathbf{o}} \circ R_{\mathbf{x}\rightarrow \mathbf{i}}$:
$$\mathbf{x} \rightarrow \mathbf{o} \in R_{\mathbf{x} \rightarrow \mathbf{o}} \iff \exists \mathbf{i} \; \mathit{s.t.} \; \mathbf{i} \rightarrow \mathbf{o} \in R_{\mathbf{i} \rightarrow \mathbf{o}} \wedge \mathbf{x} \rightarrow \mathbf{i} \in R_{\mathbf{x} \rightarrow \mathbf{i}} $$
\end{definition}

\begin{figure*}[t]
    \begin{subfigure}[b]{0.75\linewidth}
  \centering
    \begin{tabular}{c}
\begin{lstlisting}[language=C++]
// Data structure definition
struct SpVec { int len; int *idx; double *val; };
// Kernel signature
void SparseDotProduct(double &v, SpVec &A, SpVec &B);
\end{lstlisting} \\
  \end{tabular}
  \caption{Data input of sparse dot product as arguments to kernel function.}
  \label{fig:datainput}
\end{subfigure}

\begin{subfigure}[b]{0.75\textwidth}
\begin{align*}
IS = \{& [pA,pB,i] | A.idx(pA) = i = B.idx(pB)  \wedge 0 \leq pA < A.len \wedge 0 \leq pB < B.len \} 
\end{align*}
\caption{Iteration space.}
\label{fig:sparse-is}
\end{subfigure}

\begin{subfigure}[t]{0.9\textwidth}
\centering
\begin{tabular}{c}
\begin{lstlisting}[language=C++,escapechar=|]
for (int pA = 0; pA < A.len; ++pA)|\label{line:a-loop0}|
 i = A.idx[pA]; // i-loop degenerates into assignment
 for (int pB = 0; pB < B.len; ++pB)
  if (i == B.idx[pB])|\label{line:s-is-start}|
   v += A.val[pA] * B.val[pB];|\label{line:s-is-end}|
\end{lstlisting}
\end{tabular}
\caption{Dot product between two sparse vectors resulting from polyhedral scanning in SPF
.}
\label{fig:sparse-dot}
\end{subfigure}

\begin{subfigure}[t]{0.8\textwidth}
\centering
\begin{tabular}{c}
\begin{lstlisting}[language=C++,escapechar=|]
pB = 0;
for (int pA = 0; pA < A.len; ++pA) {
 i = A.idx[pA];
 while (pB < B.len && i > B.idx[pB]) ++pB;
 if (pB < B.len && i == B.idx[pB])
   { v += A.val[pA] * B.val[pB]; ++pB; }
}
\end{lstlisting}
\end{tabular}
\caption{Example optimized code generated by our framework.}
\label{fig:sparse-dot-opt}
\end{subfigure}

\caption{Dot product between sparse vectors.}

\label{fig:sparse-dot-all}
\end{figure*}

\begin{figure}[t]
\centering
\includegraphics[width=0.65\textwidth]{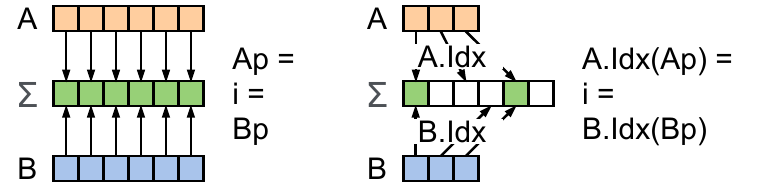}
\caption{Iteration space comparison between dense and sparse dot product. Coordinate index $i$ is introduced to illustrate how nonzeros of the same coordinates are matched to produce result.}
\label{fig:sparse-illustration}
\end{figure}

\subsection{Sparse Polyhedral Framework}

Polyhedral frameworks 
cannot directly represent sparse tensor computations, which exhibit non-affine subscript expressions and loop bounds. Fig.~\ref{fig:sparse-dot-all} illustrates 
the dot product using two sparse vectors.  Fig.~\ref{fig:sparse-illustration} shows the difference between a dense version, which computes the sum of pairwise products of all elements of two vectors, and the version that uses sparse vectors, where products are only computed when the corresponding element of both vectors is nonzero.  
Fig.~\ref{fig:datainput} represents the \emph{layout} of the input sparse vectors using the \texttt{\textcolor{blue}{struct} SpVec}. The nonzero values are stored in \texttt{A.Val} and their coordinates are in \texttt{A.idx}.
Because \texttt{A.idx} is used to encode coordinates of the nonzero values, it is commonly referred to as an \emph{index array}.  Accesses through \texttt{A.idx} introduce indirection and unknown bounds and conditions, and are not in the affine domain.

To represent this computation, the Sparse Polyhedral Framework (SPF) 
introduces \emph{uninterpreted functions} (UFs) in Presburger formulae to represent runtime values of index array references and other non-affine indices and loop bounds~\cite{Strout18}. In SPF, uninterpreted functions $A.idx$ and $B.idx$ are used to describe the combined iteration space in Fig.~\ref{fig:sparse-is}.  In Fig.~\ref{fig:sparse-dot}, we can scan the points in this iteration space and use the condition at line 3 to ensure that both vector elements are nonzero before adding their product to the sum.  The resulting code   
\emph{co-iterates} over the common nonzero elements in the vectors.  It requires a full sweep over $B$ for every element of $A$ so the time complexity of this code is $O(A.len * B.len)$. 

\begin{definition}
(Uninterpreted function (UF)) An uninterpreted function $f$ with arity of $m$, represents a mapping of $\mathbb{Z}^m \rightarrow \mathbb{Z}$.
\end{definition}

In this work, UFs are used in Presburger formulae as another term type in the affine constraints, along with index variables. We allow arguments to the UFs
to be an 
affine combination of integer set variables, integer constants, and other UFs. This ability enables polyhedral analysis to be performed on the UFs' arguments that can deduce (in-)equality relations on the arguments. 

\begin{figure*}[t!]
    \centering
    \includegraphics[width=0.8\textwidth]{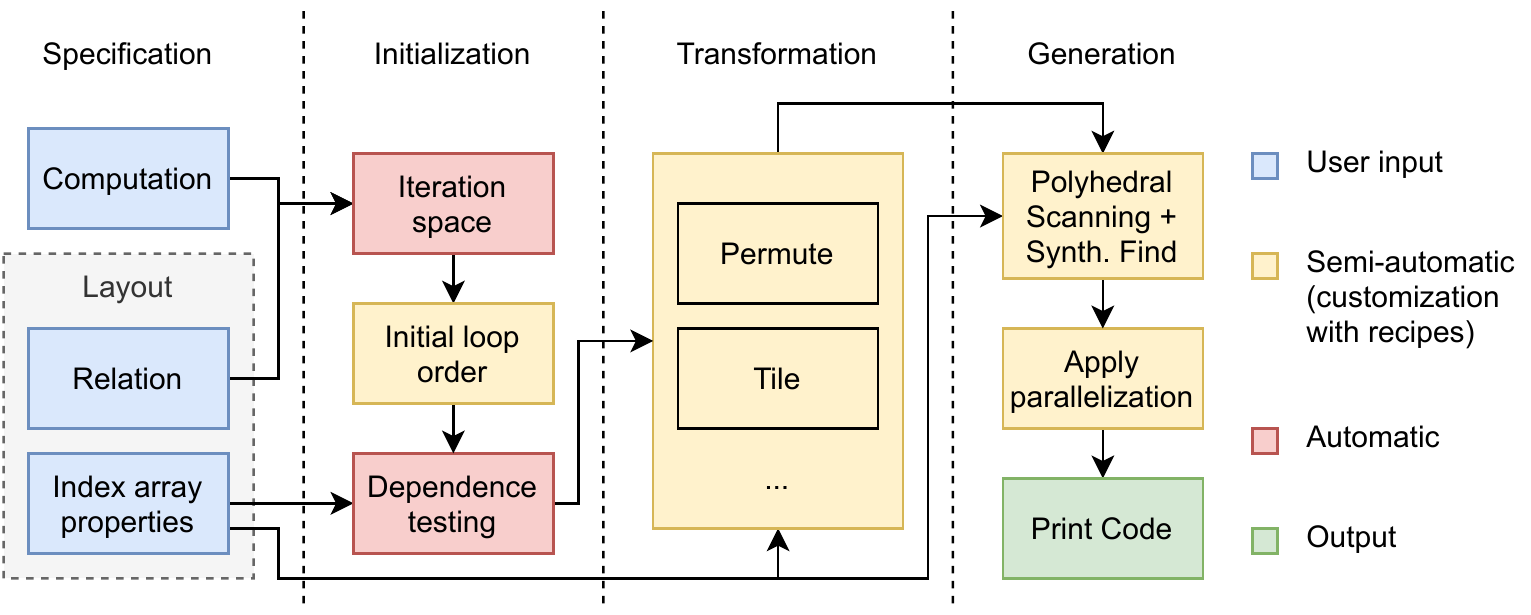}
    \caption{Overview of the code generation process for sparse tensor contraction in our polyhedral framework.}
    \label{fig:process}
\end{figure*}

\subsection{Overview of Approach}
In this paper, we optimize co-iteration using extensions to the Sparse Polyhedral Framework, thus producing
the optimized code in 
Fig.~\ref{fig:sparse-dot-opt} with time complexity $O(A.len + B.len)$. 
There are two central ideas in this approach: 
1) we compose the logical iteration space of the computation with the iteration space over the layout of the sparse tensors; and, 2)  
the conditions of the co-iteration are derived using polyhedral scanning to iterate over the elements of one sparse tensor and look up corresponding elements in the other sparse tensors using 
a \texttt{find} algorithm. In Fig.~\ref{fig:sparse-dot-opt}, \texttt{find} is implemented using the \texttt{while} loop and conditional, which represents a sequential iteration over ordered tensors that store their nonzeros in increasing coordinate order. Whether a find algorithm can be used is determined 
during code generation using an SMT solver, with the set of constraints arising from the layout specification.

Fig.~\ref{fig:process} illustrates the four stages of this approach. We begin with a specification of 
both the computation and layout, as described in Section~\ref{sec:layout}. The second stage derives the iteration space of the computation, described in Section~\ref{sec:combine}. In the third stage, we apply polyhedral transformations, as described in Section~\ref{sec:ufprops}.  The last stage generates the code using the polyhedral scanning of transformed iteration space combined with code synthesis of \texttt{find} using an SMT solver, as described in 
Section~\ref{sec:inversion}. The resulting code is then parallelized using OpenMP pragmas.
\section{Layout as Physical-to-Logical Relation}
\label{sec:layout}

In this section, we describe how sparse layouts can be represented in a sparse polyhedral framework. The key focus of this paper is to show that such a description can be incorporated into automated code generation of 
sparse tensor contraction.  In practice, layouts can be described by compiler developers, library writers, or expert programmers, where end users need not be directly exposed to these descriptions.  

A \emph{layout} is a \emph{physical} ordering of the data in memory. 
Typically, a layout represents the nonzero tensor values and a collection of auxiliary \emph{index arrays} that record coordinate information for the nonzeros so as to 
preserve the underlying \emph{logical} view of the data. 
We define a relation 
$R_{\mathbf{p} \rightarrow \mathbf{g}}$ that maps nonzero elements in the sparse tensor representing the 
physical space $\mathbf{p}$ to their corresponding logical coordinates $\mathbf{g}$.  
In $R_{\mathbf{p}\rightarrow \mathbf{g}}$, index arrays are represented by uninterpreted functions by definition because they contain read-only runtime values accessed through integer indices (arguments).

\begin{table}[b]
    \centering
        \caption{Properties of uninterpreted function $f$ representing an index array that has one dimension $a$. $\bowtie$ represents any order comparison conditional. Examples of these properties are present in the layouts described by Table~\ref{tab:layout-defs}, used in the experiments. Note that universal quantification, $\forall$, is assumed for $\mathbf{i}$, $\mathbf{i'}$, $a$, $a'$.}
    \begin{tabular}{l|c|c}
        \textbf{Array property} & \textbf{Table~\ref{tab:layout-defs} Layout} & \textbf{Definition} \\ \hline \hline
        Range & BCSR & $MIN <= f \wedge f <= MAX$ \\ \hline
        Injectivity & Unsorted-COO & $a \neq a' \rightarrow f \neq f' $ \\\hline
        (Strict) Monotonicity & SV/DCSR/Sorted-COO & $a \bowtie a' \rightarrow f  \bowtie f'$ \\ \hline
        Periodic Monotonicity & CSR/DCSR/BCSR & $period(\mathbf{i}) \wedge a \bowtie a' \rightarrow f  \bowtie f'$ \\  \hline
        Co-vary (w.$\mathit{g}$) Monotonicity & Sorted-COO & $g(\mathbf{i}) = g(\mathbf{i'}) \wedge a \bowtie a' \rightarrow f  \bowtie f'$ \\ \hline
    \end{tabular}

    \label{tab:arrayprop}
\end{table}

While values of index arrays cannot be determined until runtime, properties associated with their values are sometimes statically known and are useful to static optimization. Table~\ref{tab:arrayprop} lists 
simplified versions of the index array properties we use, demonstrated with a single argument $a$. These properties are expressed using logical formulae with guard conditions on array indices and constraints on array values as in~\citet{whats,mahdiarrayprop}.
The arguments to the uninterpreted functions can be 
affine combinations of constants, indices, and uninterpreted functions. Additionally, if there are non-affine expressions in the relation, they too can be modeled as uninterpreted functions. Compared to the goal-oriented uses of index array properties in prior works, such as for disproving dependences~\cite{mahdiarrayprop}, index array properties in this work are a component of the layout's description that targets the general question of code generation for sparse computation.

\begin{figure}[t]
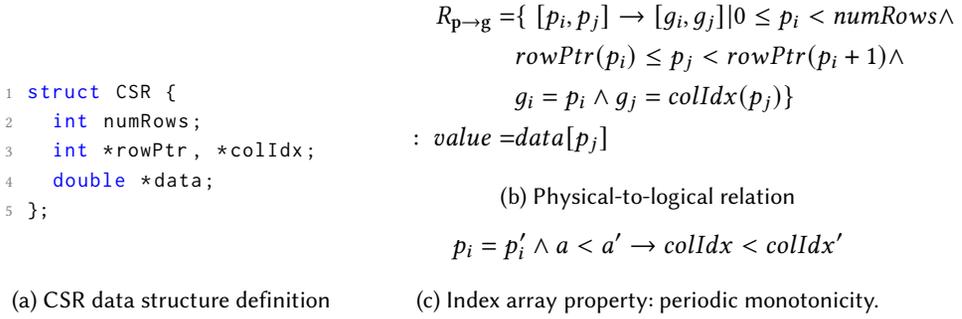

  \centering
\begin{subfigure}[b]{0.45\textwidth}
\centering
\begin{tabular}{c}
\begin{lstlisting}[language=C++,escapechar=|]
struct CSR {
  int numRows;
  int *rowPtr, *colIdx;
  double *data;
};
\end{lstlisting}\\[40pt]
\end{tabular}
\caption{CSR data structure definition}
\label{fig:csr-def}
\end{subfigure}
~
\begin{subfigure}[b]{0.45\textwidth}
\begin{align}
  R_{\mathbf{p} \rightarrow \mathbf{g}} = & \{\; [p_i,p_j] \rightarrow [g_i,g_j] | 0 \leq p_i < numRows \wedge  &\nonumber \\
  &rowPtr(p_i) \leq p_j < rowPtr(p_i + 1) \wedge  & \nonumber  \\
  & g_i = p_i \wedge g_j = colIdx(p_j) \} & \nonumber \\ 
   : \; value = & data[p_j]  & \nonumber 
\end{align}
\caption{Physical-to-logical relation}
\label{fig:csr-layout-pg}
\vspace{-1em}
\begin{equation}
    p_i = p_i' \wedge a < a' \rightarrow colIdx < colIdx' \nonumber
    \label{eq:csrProps}
\end{equation}
\caption{Index array property: periodic monotonicity.}
\label{fig:csr-layout}
\end{subfigure}
\caption{Compressed Sparse Row (CSR) layout specification}
    \label{fig:csr}
\end{figure}
\begin{figure}[t]
\centering
\vspace{-1em}
\begin{tabular}{c}
\begin{lstlisting}[language=C++]
for (int p_i = 0; p_i < csr.numRows; ++p_i)
 for (int p_j = csr.rowPtr[p_i]; p_j < csr.rowPtr[p_i+1]; ++p_j) {
  g_i = p_i;  g_j = csr.colIdx[p_j];  value = csr.data[p_j];
  y[g_i] = value * x[g_j]; |\label{line:csr-spmv}|}
\end{lstlisting}
\end{tabular}
\caption{Generated code that iterates over nonzeros in CSR and uses them to compute sparse matrix vector multiplication, $y(i) = A(i,j) * x(j)$. Indices $g_i$ and $g_j$ represent the logical indices associated with the $value$. The computation on Line~\ref{line:csr-spmv} demonstrates how these logical indices can be used.}
\vspace{-1em}
\label{fig:csrIter}
\end{figure}

Fig.~\ref{fig:csr-layout-pg} presents $R_{\mathbf{p} \rightarrow \mathbf{g}}$ for the common \emph{Compressed Sparse Row} (CSR) layout declared as in Fig.~\ref{fig:csr-def}.  Index array \texttt{rowPtr} refers to the first nonzero element of each row in the \texttt{val} vector of nonzeros.  Index array \texttt{colIdx} refers to the column associated with each nonzero element.  Note that $R_{\mathbf{p}\rightarrow \mathbf{g}}$ describes both the layout indices $p_i, p_j$ for iterating over the sparse layout, and the logical indices $g_i, g_j$ for iterating over a 2-d tensor $A$.  As nonzero elements in CSR layout are sorted by row, a periodic monotonicity property exists for \texttt{colIdx} array, which is expressed by the logical formula in Fig.~\ref{fig:csr-layout}. This logical formula denotes when two iteration instances containing $colIdx$ are induced from the same physical index $p_i=p_i'$. If the first instance's argument, $a$, is smaller than that of the second, $a'$, then the first value, $colIdx$, will also be smaller than the second, $colIdx'$. 
Applying polyhedra scanning to this description, our compiler can generate the code in Fig.~\ref{fig:csrIter}, which iterates over all nonzeros in the layout.  

Index arrays aid in providing coordinates corresponding to nonzeros in a sparse tensor, so the layout description can be specified as a mathematical relation from the data structure scalar and array fields to the tensor. 
By expressing layouts as 
polyhedral relations, tensor contraction operations with sparse tensors can be composed using polyhedral set and relation operations, as discussed in the following section. 

\section{Deriving Iteration Space from Computation and Layout}
\label{sec:combine}

A sparse tensor contraction computation may involve two or more sparse tensors, potentially using different layouts. 
This section describes how such layouts, specified using sparse polyhedral relations, can be combined with the access pattern in the computation to derive 
an iteration space.
This enables subsequent 
transformations to the iteration space.

As an illustrative example, consider the tensor contraction matrix multiplication. 
In tensor index notation, this contraction is written as $C(i,j) = A(i,k) * B(k,j)$. 
The following iteration space results for loop indices $\mathbf{I} = [i,k,j]$:
$$\mathit{IS}_{\mathbf{I}} = \{[i,k,j] : 0 \le i<N \wedge 0\le j<N \wedge 0\le k<N\}$$
We represent an access expression for tensors $C$, $A$ and $B$ 
as a mapping from the computation's iteration space $\mathbf{i}$ to the tensor's data space.  
\begin{align*}
A^{(C)}_{\mathbf{I} \rightarrow \mathbf{g}} &= \{[i,k,j] \rightarrow [g_i,g_j] | g_i = i \wedge g_j = j \}\\
A^{(A)}_{\mathbf{I} \rightarrow \mathbf{g}} &= \{[i,k,j] \rightarrow [g_i,g_j] | g_i = i \wedge g_j = k \}\\
A^{(B)}_{\mathbf{I} \rightarrow \mathbf{g}} &= \{[i,k,j] \rightarrow [g_i,g_j] | g_i = k \wedge g_j = j \}
\end{align*}

To determine the 
part of logical iteration space $\mathbf{I}$ that accesses nonzeros in the sparse tensor representation, 
we derive $Q_{\mathbf{p} \rightarrow \mathbf{I}}$: the composition of the layout description $R_{\mathbf{p} \rightarrow \mathbf{g}}$ with the access mapping $A_{\mathbf{I} \rightarrow \mathbf{g}}$.
Using CSR layout for A as described in Fig.~\ref{fig:csr-layout}, we have the relation as follows:
\begin{align*}
& (A^{(A)}_{\mathbf{I} \rightarrow \mathbf{g}})^{-1} \circ R^{(A)}_{\mathbf{p} \rightarrow \mathbf{g}}   = \\
Q^{(A)}_{\mathbf{p} \rightarrow \mathbf{I}}  = & \{\; [p_i,p_j] \rightarrow [i,k,j] | 0 \leq p_i < numRows \wedge \\
  & rowPtr(p_i) \leq p_j < rowPtr(p_i + 1) \wedge  \nonumber  \\
   & i = p_i \wedge k = colIdx(p_j) \}
\end{align*}


\begin{definition}
Range of a Presburger relation, 
$R_{\mathbf{o}} = \mathit{Range}(R_{\mathbf{i} \rightarrow \mathbf{o}})$:
$$\mathbf{o} \in R_{\mathbf{o}} \iff \exists i \; \mathit{s.t.} \; \mathbf{i} \rightarrow \mathbf{o} \in R_{\mathbf{i} \rightarrow \mathbf{o}}$$\label{def:range}
\vspace{-1em}
\end{definition}

$\mathit{Range}(Q^{(A)}_{\mathbf{p} \rightarrow \mathbf{I}})$ corresponds to all positions in the iteration space of $\mathbf{i}$ that have a nonzero or explicit zero stored in the sparse format of $A$.

The new iteration space accessing multiple sparse tensor layouts will be the intersection or union of the parts of the original iteration space that accesses sparse tensors based on whether the computation is a multiplication (intersection) or an addition (union). This is as a sparse polyhedral 
definition of merge lattices proposed by TACO~\cite{taco,HentrySparseArray2021}.
When multiple tensors are multiplied, such as $A(i,k) * B(k,j)$ in the example, the value may be nonzero if and only if both sparse tensors store the value for the iteration $(i,k,j)$. Thus, intersection is used to combine layouts relations under the logical access of the computation in multiplication, $P$ for product.
\begin{align}
P &= \mathit{Range}(Q_{\mathbf{p}\rightarrow \mathbf{I}}^{(A)}) \cap \mathit{Range}(Q_{\mathbf{p}\rightarrow \mathbf{I}}^{(B)}) \nonumber
\end{align}

The iteration space for transformation and code generation also includes the layout for the output tensor and the dense iteration space. In our implementation, we support dense output tensors as well as sparse tensors with known sparsity. Under such cases,
the output of the tensor contraction can be treated as another product term and intersected with $P$. Known output sparsity is common in core computations of data analytics~\cite{10.1145/2487575.2487677} and graph neural network (GNN)~\cite{DBLP:journals/corr/abs-1909-01315} such as Sampled Dense-Dense Matrix Multiplication (SDDMM) and Sparse Matrix Times Dense Vector (SpMM), and can be generated for other sparse computations using inspectors. The dense iteration space represent optional bounds that can bound the computation to a sub-region of the valid iterations specified by the layouts, such as when a computation only operates on the lower triangular region of a layout. Further bound by the dense iteration space, we have the iteration space for polyhedral transformation and code generation.
$$\mathit{IS} = P \cap \mathit{Range}(Q_{\mathbf{p}\rightarrow \mathbf{I}}^{(C)}) \cap \mathit{IS}_{\mathbf{I}}$$

Note that the existential operation on the input indices in the definition for the range operations (Definition~\ref{def:range}) does not guarantee the input indices can be eliminated through simplification. In fact, with $Q_{\mathbf{p} \rightarrow \mathbf{I}}^{(A)}$, the input indices hold special meaning in the iteration space and may reference index arrays. Some of these indices not only have to be "rematerialized" during the code generation but can also be involved in transformations like positional tiling~\cite{Senanayake2020}. 

Instead of relying on the code generator to make decisions when to rematerialize existential variables, we pull these indices out of the existential operations and make them part of the set variables of the iteration space. Thus, the iteration space will consist of all layout indices and the indices of the computation.
\begin{align*}
    \mathit{IS} & = P \cap \mathit{Range}(Q_{\mathbf{p}\rightarrow \mathbf{I}}^{(C)}) \cap \mathit{IS}_{\mathbf{I}} \\
    &= \mathit{Range}(Q_{\mathbf{p}\rightarrow \mathbf{I}}^{(A)}) \cap \mathit{Range}(Q_{\mathbf{p}\rightarrow \mathbf{I}}^{(B)}) \cap \mathit{Range}(Q_{\mathbf{p}\rightarrow \mathbf{I}}^{(C)}) \cap \mathit{IS}_{\mathbf{I}} \\
    & = \{ [\mathbf{I}] | (\exists \mathbf{p}^{(A)} \ldots) \wedge (\exists \mathbf{p}^{(B)} \ldots) \wedge (\exists \mathbf{p}^{(C)} \ldots) \wedge \ldots \} \\ 
    & = \{ [\mathbf{I}] | (\exists \mathbf{p}^{(A)} (\exists \mathbf{p}^{(B)} (\exists \mathbf{p}^{(C)} \ldots \wedge \ldots \wedge \ldots \wedge \ldots))) \} \\
    & = \{ [\mathbf{p}^{(A)}, \mathbf{p}^{(B)}, \mathbf{p}^{(C)}, \mathbf{I}] | \ldots \wedge \ldots \wedge \ldots \wedge \ldots \}
\end{align*}

For additions, union can be used, such as, for $v() = A(i) + B(i)$, $S$ for sum, $S = \mathit{Range}(Q_{\mathbf{p}\rightarrow \mathbf{I}}^{(A)}) \cup \mathit{Range}(Q_{\mathbf{p}\rightarrow \mathbf{I}}^{(B)})$. However, union will cause implicit zeros of one of the tensors being accessed when some other tensors are not zero. Alternatively, the polyhedral framework can use statement splitting to give each addition term its own iteration space to guard their execution. For example $v() = A(i) + B(i)$ can be split into $v() = A(i)$ and $v() = B(i)$. Either approach benefits from additional optimizations to save space or fuse separate loops, outside the scope of this paper.

\section{Polyhedral Analysis \& Transformations}

\label{sec:ufprops}

In the previous section, we demonstrated how to combine 
relations defined by the layouts and the 
iteration space of the computation to form the iteration space of the generated code. 
In this section, we show how using a sparse polyhedral representation permits reasoning about parallelism and composing code transformations.  These concepts are exemplified with discussions on dependence testing and tiling.   

\begin{figure}[t!]
    \centering
\begin{subfigure}[b]{0.4\textwidth}
\centering
\begin{tabular}{c}
\begin{lstlisting}[language=C++,escapechar=|]
struct DCSR {
    int numRows;
    int *rowIdx, *rowPtr;
    int *colIdx;
    double *data;
};
\end{lstlisting}\\[40pt]
\end{tabular}
\caption{Data structure definition.}
\label{fig:dcsr-def}
\end{subfigure}
\begin{subfigure}[b]{0.5\textwidth}
\begin{align*}
R_{\mathbf{p} \rightarrow \mathbf{g}} &= \{\; [p_i,p_j] \rightarrow [g_i, g_j] | 0 \leq p_i < numRows \wedge  \nonumber \\
& rowPtr(p_i) \leq p_j < rowPtr(p_i + 1) \wedge   \nonumber  \\
& g_i = rowIdx(p_i) \wedge g_j = colIdx(p_j)\}  \nonumber \\ 
   : \; value &= data[p_j]  
\end{align*}
\begin{align*}
a < a' &\rightarrow rowIdx < rowIdx' \\
p_i = p_i' \wedge a < a' &\rightarrow colIdx < colIdx'  
\end{align*}
\caption{Layout definition.}
\label{fig:dcsr-layout}
\end{subfigure}

\begin{subfigure}[t]{0.9\textwidth}
\centering
\begin{tabular}{c}
\begin{lstlisting}[language=C++,escapechar=|]
void spmv(double *y, DCSC A, double *x) {
 for (int p_i = 0;p_i < A.numRows; ++p_i)
  for (int p_j = A.rowPtr[p_i]; j < A.rowPtr[p_i+1]; ++p_j)
   y[A.rowIdx[p_i]] += A.data[p_j] * x[A.colIdx[p_j]];}
\end{lstlisting}
\end{tabular}
\caption{Sparse matrix vector multiplication, $y(i) = A(i,j) * x(j)$.}
\label{fig:dcsr-spmv}
\end{subfigure}
    \caption{Doubly compressed sparse row (DCSR).}
    \label{fig:dcsr}
    \vspace{-1em}
\end{figure}

\subsection{Dependence Testing}
In general, dependence testing determines if it is safe to parallelize a loop or apply a transformation, 
realized with a \emph{dependence polyhedron}~\cite{feautrier1991dataflow,violateddependence} in polyhedral frameworks.
Tensor contraction expressions exhibit specific data dependence patterns.
In matrix vector multiplication, $y(i) = A(i,j) * x(j)$, $j$ is a contraction index that carries reduction dependences since $y(i)$ is the sum over $A(i,j)*x(j)$ products;
$i$ is a free index without loop-carried dependences.  
Since the generated code also iterates over the layout indices, the compiler must  translate dependence relations to refer to layout indices, which can be described in the sparse polyhedral framework.

 We derive the dependence polyhedron for two accesses to 
the same tensor, where one of them is a write, using the combined iteration space of Section~\ref{sec:combine}, and the lexicographical loop
 order.  Because the sparse layout will not change the inherent dependences
of the computation, we observe dependences from the logical access, allowing us to circumvent complexities arises from performing dependence analysis on the 
value arrays through indirect accesses with index arrays in the sparse layout. This dependence polyhedron can be also combined with the dense dependences to determine whether the original order of the computation is preserved.

Index array properties added to the dependence polyhedron allow dependence testing to be more precise, 
as shown with the \emph{Doubly Compressed Sparse Row} (DCSR) layout 
in Fig.~\ref{fig:dcsr}.
Consider the \texttt{p\_i} loop in Fig.~\ref{fig:dcsr-spmv}; the information 
that $A.rowIdx$ is monotonically increasing proves that loop \texttt{p\_i} does not carry a dependence.

\subsection{Tiling}

We can express transformations such as tiling as relations on the iteration space, such as in the example of 
Fig.~\ref{fig:tiling}. The combined iteration space of Section~\ref{sec:combine} guards execution in Fig.~\ref{fig:tiling}(a) based on the value of a UF $f$ representing an index array.  
Tiling transforms the loop into loops on Line~1 and 3 of  Fig.~\ref{fig:tiling-opt}. However, the compiler introduces the condition on Line~2 due to the monotonicity of $f$ as an optimization after tiling is applied. This introduced condition will significantly reduce the number of tiles 
executed and improves the performance.

\begin{figure}[t]
    \centering
\begin{subfigure}[c]{0.45\textwidth}
\centering
\begin{tabular}{c}
\begin{lstlisting}
for (i = 0; i < n; ++i)
 if (f[i] == k) S0();
\end{lstlisting}
\end{tabular}
\caption{Original program where $k$ is a loop invariant value in $i$-loop.}
\label{fig:tiling-orig}
\end{subfigure}
\begin{subfigure}[c]{0.45\textwidth}
\begin{align*}
a < a' &\rightarrow f < f'
\end{align*}
\caption{Monotonicity of $f$.}
\label{fig:tiling-prop}
\end{subfigure}

\begin{subfigure}[t]{0.6\textwidth}
\centering
\begin{tabular}{c}
\begin{lstlisting}
for (ti = 0; 8*ti < n; ++ti)
 if (f[8*ti]<=k && f[min(n, 8*ti + 8) - 1]>=k)
  for (i = 8*ti; i < min(n, 8*ti + 8); ++i)
   if (f[i] == k) S0();
\end{lstlisting}
\end{tabular}
\caption{Tiled computation with induction loop $ti$. Line 2 demonstrate the constraints introduced by monotonicity of $f$.}
\label{fig:tiling-opt}
\end{subfigure}
    \caption{Tiling a sparse computation with UF properties.}
    \label{fig:tiling}
\end{figure}


To achieve sparse tiling, these constraints are generated from matching the UF property expression with bounds produced from the polyhedral scanning.

\section{Code Generation}

\label{sec:inversion}


With the combined iteration space in Section~\ref{sec:combine}, we can generate code that iterates over the parts described by the layout indices --- subsections, rows, columns, or nonzeros --- of one or more sparse tensors, and look up corresponding parts in the next sparse tensor in its layout indices using \emph{find} algorithms, as described in this section.


\subsection{Co-iteration using Polyhedral Scanning}


Polyhedral scanning can generate loops that handle co-iteration with if conditions from the iteration space constraints, such as $IS$ in Figure~\ref{fig:sparse-is} for the sparse-sparse vector dot product. Such code is generated through classic polyhedral scanning algorithms~\cite{chill}, where conditions involving each loop index are produced through set operations. Loops are generated from these conditions by extracting the lower bounds and upper bounds from the index conditions. Conditions other than bounds are turned into stride when they specify modulo equality and if conditions if otherwise. For example, the polyhedral scanning produces the following constraint for the $pB$ loop.
\begin{equation}
    0 \leq pB \wedge pB < B.len \wedge i = B.idx(pB)
    \label{eq:pBCond}
\end{equation}
Note that in this relation, the first two terms specify the loop bound, and the third term specifies the if condition.

The resulting code is shown in Fig.~\ref{fig:sparse-dot}, with  
two loops, one for each layout --- $pA$ and $pB$ --- and a condition in the $pB$ loop that relates locations in these two sparse layouts.
This code will work regardless of whether the elements of either vector are sorted.  However, more efficient code can be generated when it is known that the vectors are sorted, discussed next.

\begin{figure}[t]
\begin{subfigure}[b]{0.495\textwidth}
    \centering
\begin{tabular}{l}
\begin{lstlisting}[language=C++,escapechar=|]
|\colorbox{brown!20}{pB=0;}|

for (int pA=0;pA<A.len;++pA) {
 i = A.idx[pA];
 |\colorbox{green!20}{\textcolor{blue}{while}(pB<B.len\&\&i>B.idx[pB]) ++pB;}|
 |\colorbox{green!20}{\textcolor{blue}{if}(pB<B.len\&\&i==B.idx[pB])}|
  { v+=A.val[pA]*B.val[pB];|\colorbox{green!20}{++pB;}| }}
\end{lstlisting}\\[43pt]
\end{tabular}
    \caption{Sequential iteration, \texttt{SeqIter}.}
    \label{fig:sequentialiteration}
    \end{subfigure}
\begin{subfigure}[b]{0.43\textwidth}
    \centering
\begin{tabular}{r}
\begin{lstlisting}[language=C++,escapechar=|]
|\colorbox{brown!20}{\textcolor{blue}{for} (int pB=0;pB<pB.len;++pB)}|
  |\colorbox{brown!20}{hashB[B.idx[pB]] = pB;}|
for (int pA=0;pA<A.len;++pA) {
  i = A.idx[pA];
  |\colorbox{green!20}{pB = hashB.find(i)}|
  |\colorbox{green!20}{\textcolor{blue}{if} (pB!=hashB.notfound)}|
    v+=A.val[pA]*B.val[pB];}
\end{lstlisting}
\end{tabular}
    \caption{\texttt{HashMap}.}
    \label{fig:hashmap}
\end{subfigure}
    \caption{Find algorithms and sparse vector dot product. Code belonging to the templates are highlighted.}
    \label{fig:algorithms}
\end{figure}

\subsection{Optimized Co-Iteration using Synthesis of Find}

The conditions specified in \ref{eq:pBCond} that produce the $pB$ loop at line 3 and if condition at line 4 in Figure~\ref{fig:sparse-dot} can be alternatively described as a find: looking for index $pB$ within the loop bound such that $B.idx(pB) = i$. When it is treated as a \texttt{find(B,pB)}, different find algorithms can be used to replace this loop. We illustrate two examples of find algorithms that replace the $pB$ loop in Fig~\ref{fig:algorithms}, \texttt{SeqIter}, which refers to sequential iteration, and \texttt{HashMap}. \texttt{SeqIter} find matches by scanning through the loop range of $pB$ using an inequality version of the find condition until a match is found or no matches are possible. While the scanning is linear, the initialization of $pB$ affects if scanning resumes from the last saved position. Through amortization, \texttt{SeqIter} is of complexity $O(A.len + B.len)$. \texttt{HashMap} uses a hashmap to perform the find. It can only handle equality find conditions and has a complexity of $O(A.len + B.len)$, including the initialization cost of the hashmap.

Each find algorithm has a basic skeleton of code associated with it. This skeleton is captured in the code generator with a template for each algorithm. 
During code generation, the find algorithm's template will be filled in with constraints arising from the computation, index array properties, and loop permutation order: 
they can be integer values or algebraic expressions. 
Assumptions of each algorithm determine which templates are valid and how to generate the template arguments. For example, the \texttt{SeqIter} code in Fig.~\ref{fig:sequentialiteration} is only valid when the elements of each vector are sorted. For a given find algorithm, its assumptions are encoded using logical formulae.  The generation of the find algorithm will try to match the assumptions with conditions from iteration spaces, index array properties, and generated template arguments. 

We use a \emph{satisfiability modulo theory} (\emph{SMT}) solver to prove if these assumptions are met. Template arguments are first generated through enumeration or construction, and then the assumptions are checked with the generated arguments.
This method of generating code segments by proving generated code with an MT solver or high-order logic (HOL) provers are commonly referred to as \emph{code synthesis}\cite{lezama2006sketching,synthesis}. The details of this synthesis process are presented in auxiliary material for this paper.

\begin{algorithm}[t]
\SetAlgoVlined
\KwIn{$\mathit{IS}$: Extended Iteration Space. $\mathit{Idx}$: Loop indices in combined iteration space $IS$ from outermost to innermost.}
 \For {Index $i\in \mathit{Idx}$}{
   $L_i = $ scan iteration ranges of $i$ in $\mathit{IS}$
   \tcc*{Polyhedral}
   $\mathit{UF} = $ uninterpreted functions applied with $i$ as the last unbound index\; \label{line:identify}
   $\mathit{C_{i}} = \mathit{map}\langle UF,bounds \rangle$\;
   \For{$\mathit{uf} \in \mathit{UF}$ \label{line:extract}}{
     \uIf(\tcc*[f]{Polyhedral}){Scan equality/range $R$ of $\mathit{uf}$ in $\mathit{IS}$}{ 
       Insert $(\mathit{uf}, R)$ in $\mathit{\mathit{C_{i}}}$\;
     }
     \label{line:extract-end}
   }
   \uIf(\tcc*[f]{SMT}){Find algorithm $A$ can be applied on $L_i,\mathit{\mathit{C_{i}}}$ \label{line:check}}{ 
     Generate $A$\; \label{line:generate}
   }\uElse{
     Generate \texttt{for} loop with bound $L_i$ and \texttt{if} condition $\mathit{\mathit{C_{i}}}$\;
   }
 }
 \caption{Augmented polyhedral scanning.}
 \label{alg:codegen}
\end{algorithm}

\subsection{Code Generation Algorithm}
We present the tensor contraction code generation algorithm in  
Algorithm~\ref{alg:codegen}, which augments polyhedral scanning to leverage an SMT solver to synthesize find algorithms.
Each loop index in the combined iteration space $IS$ from Section~\ref{sec:combine} is processed from outermost to innermost.
Line~\ref{line:identify} identifies all uninterpreted functions that are fully bound at this loop level. Line~\ref{line:extract}-\ref{line:extract-end} extracts find conditions produced by the combined iteration space. Line~\ref{line:check}-\ref{line:generate} uses the SMT solver to detect and generate find algorithm $A$ as an alternative to the loop and if conditions.

\begin{table*}[t]
    \centering
    \begin{tabular}{c|c}\hline
        \textbf{Data structure definition} & \textbf{Layout definition} \\ \hline \hline
         \multicolumn{2}{c}{Sparse Vector (SV) [C(unique)]} \\ \hline
         \begin{lstlisting}[language=C++,escapechar=|]
struct SpVec {
    int len;
    int *idx;
    double *val;
};
\end{lstlisting} &
          \makecell{$\!\begin{aligned}
R_{\mathbf{p};\mathbf{g}} = & \{\; [p_i] \rightarrow [g_i] | 0 \leq p_i < len \wedge g_i = idx(p_i)\} \nonumber \\ 
   : \; value = & val[p_i]
\end{aligned}$\\
$\!\begin{aligned}
a < a' \rightarrow& idx \leq idx'
\end{aligned}$}
         \\ \hline \hline
         \multicolumn{2}{c}{Compressed Sparse Row (CSR)~\cite{saad2003} [D,C(unique)]: Fig.~\ref{fig:csr}} \\ \hline \hline
           \multicolumn{2}{c}{Doubly-compressed Sparse Row (DCSR)~\cite{buluc2008} [C(unique),C(unique)]: Fig.~\ref{fig:dcsr}} \\  \hline \hline
        \multicolumn{2}{c}{Coordinate (COO)~\cite{saad2003} [C(not-unique),Q]} \\ \hline
        \begin{lstlisting}[language=C++,escapechar=|]
struct COO {
    int numNNZ;
    int *rowIdx;
    int *colIdx;
    double *data;
};
\end{lstlisting} & \makecell{$\!\begin{aligned}
R_{\mathbf{p};\mathbf{g}} = & \{\; [p_i] \rightarrow [g_i, g_j] | 0 \leq p_i < numNNZ \wedge  \nonumber \\ & \;\;\; g_i = rowIdx(p_i) \wedge g_j = colIdx(p_i)\} \nonumber \\ 
   : \; value = & data[p_i]
\end{aligned}$\\
$\!\begin{aligned}
a < a' \rightarrow& rowIdx \leq rowIdx' \\
a < a' \wedge rowIdx(a) = rowIdx(a') \rightarrow& colIdx < colIdx'  
\end{aligned}$} \\ \hline \hline
    \multicolumn{2}{c}{Block Compressed Sparse Row (BCSR)~\cite{im2001}} \\ \hline
\begin{lstlisting}[language=C++,escapechar=|]
struct BCSR {
    int numRows;
    int *rowPtr;
    int *colIdx;
    double *data[8][8];
};
\end{lstlisting} & \makecell{$\!\begin{aligned}
R_{\mathbf{p}\rightarrow \mathbf{g}} = & \{\; [p_i,p_j,p_k,p_l] \rightarrow [g_i, g_j] | 0 \leq p_i < numRows \wedge  \nonumber \\ 
& \;\;\; rowPtr(p_i) \leq p_j < rowPtr(p_i+1) \wedge \nonumber \\ 
& \;\;\; 0 \leq p_k < 8 \wedge 0 \leq p_l < 8 \wedge \nonumber \\ 
 & \;\;\; g_i = p_i * 8 + p_k \wedge g_j = colIdx(p_j) + p_l \} \nonumber \\ 
   : \; value = &  data[p_j][p_k][p_l] 
\end{aligned}$\\
$\!\begin{aligned}
p_i = p_i' \wedge a < a' &\rightarrow colIdx + 8 \leq colIdx'
\end{aligned}$} \\ \hline \hline
\multicolumn{2}{c}{Lower triangular} \\ \hline
\begin{lstlisting}[language=C++,escapechar=|]
struct LowerTri {
    int numRows;
    double *data;
};
\end{lstlisting}&
\makecell{$\!\begin{aligned}
R_{\mathbf{p}\rightarrow \mathbf{g}} = & \{\; [p_i,p_j] \rightarrow [g_i, g_j] | 0 \leq p_i < numRows \wedge  \nonumber \\
& \;\;\; 0 \leq p_j \leq p_i \wedge g_i = p_i \wedge g_j = p_j \} \nonumber \\ 
   : \; value = & data[p_i * (p_i + 1) / 2 + p_j]
\end{aligned}$} \\ \hline \hline
\multicolumn{2}{c}{Warp Sparse Matrix Storage~\cite{ptx,mishra2021accelerating}} \\ \hline
\begin{lstlisting}[language=C++,escapechar=|]
struct CUDAmma16x16 {
    float data[16][8];
    Bits<512> offset;
};
\end{lstlisting} &
\makecell{$\!\begin{aligned}
R_{\mathbf{p}\rightarrow \mathbf{g}} = & \{\; [p_i,p_j] \rightarrow [g_i,g_j] | 0 \leq i < 16 \wedge 0 \leq p_j < 8 \wedge  & \\
& g_i = p_i \wedge g_j = p_j * 2 + \mathit{offset}(p_i * 32 + p_j * 4 + 1) \} & \\ 
   : \; value = & data[p_i][p_j]  &
\end{aligned}$\\
$\!\begin{aligned}
0 \leq \mathit{offset} < 2
\end{aligned}$} \\ \hline
    \end{tabular}
    \caption{Different layout definitions.  TACO's layout description are shown in square bracket when possible using the level formats, dense (D), compressed (C), and singleton (Q), with properties in the parenthesis per level.}
    \label{tab:layout-defs}
    \vspace{-1em}
\end{table*}

Applying Algorithm~\ref{alg:codegen} on the iteration space of sparse dot product in Fig.~\ref{fig:sparse-is}, polyhedral scanning is used to 
identify the range for loop $pA$. There are no uninterpreted functions at this loop level, and a \emph{for-}loop is generated.
For the next loop $pB$, $A.idx(pA)$ is loop invariant. When $A.idx$ and $B.idx$ are monotonically increasing, SMT solver can prove that find algorithms such as \emph{sequential iteration} and \emph{hashmap} can be applied to implement the find in $pB$. When \emph{sequential iteration} is applied, the code in Fig.~\ref{fig:sparse-dot-opt} is generated. Loop $i$ will be generated as an assignment from scanning, $i = A.idx(pA)$, and subsequently removed by dead code elimination due to no usage related to $i$.

\section{Demonstrations and Comparisons}

This section we demonstrate our proposed framework in two aspects: the versatility of our sparse layout specification and the adaptability of our code generation strategy.

\subsection{Sparse Tensor Layouts}
Table~\ref{tab:layout-defs} presents sparse tensor layouts as described in our framework using the 
approach in Section~\ref{sec:layout}. All but the last layout are used in the experimental evaluation.  The first column describes the layout using common terms or citations.  
Without loss of generality, higher order sparse tensor layouts such as Compressed Sparse Fiber (CSF) can be similarly specified with relations including uninterpreted functions representing index arrays and logical formulae describing index array properties.

In the table, the last column compares how these layouts are supported by the TACO compiler~\cite{taco}.
TACO uses level format and mode ordering to specify how and in what order dimensions are stored. Four 
level formats are defined: dense, compressed, sparse, and singleton. Mode ordering specifies the order in which levels are organized. Each level format can be further customized with 
properties like uniqueness and sortedness.
Two points of difference with TACO are (1) its 
coupling of logical dimensions to the physical dimensions, thus disallowing a logical dimension to be derived from multiple physical dimensions\footnote{BCSR defined in Table~\ref{tab:layout-defs} is not possible with TACO due to the relation on $g_j$. TACO can describe a more restricted version of BCSR with aligned $g_j = 8 * colIdx(p_j) + p_j$:  D,C(unique),D,D.}; 
and, (2) TACO 
views each dimension separately, thus disallowing relations such as $col \leq row$ in the lower-triangular matrix. 

Looking to the future of coarse-grained functional units such as 
the NVIDIA A100 sparse tensor core, we show how our approach 
describes the Warp Sparse Matrix Storage\cite{ptx,mishra2021accelerating}.
Our code generation can produce computation kernels on the host CPU for computations not natively supported by the tensor core without requiring layout changes or writing complex architecture-specific code.

\subsection{Comparison with Conjunctive Merge}

\label{sec:co-iteration-analysis}

This subsection compares the code generated by our approach as compared with that of the TACO compiler~\cite{taco}, using sparse dot product as an example. 
Specifically, Fig.~\ref{fig:sparse-sort-sort} revisits the code generated by 
our sequential iteration algorithm template, where the \texttt{i} loop at line 2 iterates over the nonzero elements in the layout of $A$, and the loop at line 4 along with the condition at line 6 looks for that element in $B$.   It only examines each element of $A$ once, and then searches adjacent elements in $B$ for the index in $A$, only visiting an element in $B$ twice if it matches an element in $A$.  Our approach can alternatively generate code with loop permutation, which iterates over $B$ and performs a lookup of $A$.  The code in Fig.~\ref{fig:sparse-sort-sort} will perform better when $A$ contains fewer nonzeros since each element of $A$ is only examined once, and vice versa for the permuted code.  By comparison, TACO's \emph{conjunctive merge algorithm}
(Fig.~\ref{fig:sparse-sort-sort-taco}) iterates over both sparse tensors in the while loop at line 2, and must examine an element of $A$ and an element of $B$ in each iteration, even if it was examined in the previous iteration. 
Note that all implementations 
have linear complexity $O(A.len + B.len)$, but the code generated by TACO exhibits more data movement. 

\begin{figure*}[t]
\centering
\begin{subfigure}[b]{0.48\textwidth}
\centering
\begin{tabular}{c}
\begin{lstlisting}[language=C++,escapechar=|]
pB = 0;
for (int pA=0;pA<A.len;++pA) {
 i=A.idx[pA];
 while (pB<B.len&&i>B.idx[pB])
  ++pB;
 if (pB<B.len&&i==B.idx[pB]) 
 { v+=A.val[pA]*B.val[pB]; ++pB; }}
\end{lstlisting}
\end{tabular}
\caption{Sequential iteration.}
\label{fig:sparse-sort-sort}
\end{subfigure}
~
\begin{subfigure}[b]{0.48\textwidth}
\centering
\begin{tabular}{c}
\begin{lstlisting}[language=C++,escapechar=|]
pA = 0; pB = 0;
while (pA < A.len && pB < B.len) {
 A0 = A.idx[pA]; B0 = B.idx[pB];
 i = min(A0,B0);
 if (A0 == i && B0 == i)
  v+=A.val[pA]*B.val[pB];
 pA += (int)(A0 == i);
 pB += (int)(B0 == i);}
\end{lstlisting}
\end{tabular}
\caption{Conjunctive merge from TACO.}
\label{fig:sparse-sort-sort-taco}
\end{subfigure}

\caption{Sparse vector dot products.}

\label{fig:sparse-gen}
\end{figure*}

The second difference relates to how we can handle a much greater set of index array properties than TACO. This is related to both the expressiveness of the layout description and the adaptivity of the code generation. Fig.~\ref{fig:sparse-multi} demonstrates a three-way co-iteration where each sparse vector involved has a different ordering on the nonzeros. TACO uses flags to specify index array properties. TACO thus can not express slight variations of the properties, such as decreasingly sorted used by $B$. When specific properties such as sortedness are not provided and the \emph{locate}~\footnote{Finding position from coordinate.} level-function is not defined on a level format, TACO will also fail to generate code as in the case of $C$, which can be described as a compressed level format without sortedness. In our framework, synthesis allows more adaptability regarding variations of index array properties. Different find algorithms, including the fallback loop implementation, allows us always to generate valid and efficient code under the constraints provided.

\begin{figure}[t]
\centering
\begin{tabular}{c}
\begin{lstlisting}[language=C++,escapechar=|]
hashmap hashC;
for (pC = 0; pC < C.len; ++pC) hashC[C.idx[pC]] = pC;
pB = B.len - 1;
for (int pA = 0; pA < A.len; ++pA) {       // pA
 i = A.idx[pA];
 while (pB < B.len && i > B.idx[pB]) --pB; // pB sequential iteration
 if (pB < B.len && i == B.idx[pB]) { 
  pC = hashC.find(i);                      // pC hashmap
  if (pC != hashC.notfound) v += A.val[pA] * B.val[pB] * C.val[pC]; 
  --pB;
 }}
\end{lstlisting}
\end{tabular}
\caption{Three-way co-iteration, computing $v = A(i) * B(i) * C(i)$, where $A$ has uniqueness and increasing monotonicity, $B$ has uniqueness and decreasing monotonicity, and $C$ has uniqueness but no monotonicity.}
\label{fig:sparse-multi}
\end{figure}

\section{Experiments}
\label{sec:experiment}

We have implemented a 
polyhedral compiler with the layout specification, dependence testing, and sparse polyhedral code generation 
extensions presented in this paper. 
In our 
implementation, we used functionalities provided by the CHiLL compiler~\citep{chill}, the Omega+ Library~\citep{Chenpldi2012} for integer set manipulation and scanning, and Z3~\citep{z3} for theory proving.

\subsection{Experiment Setup}

\begin{table}[t]
    \centering
        \caption{Computations used in comparison. }
    \begin{tabular}{c|c|c}\hline
        \textbf{Name} & \textbf{Notation} & \textbf{Matrix size} \\ \hline \hline
        \texttt{SpMV} & $y(i) = A(i,j) * x(j)$ & SuiteSparse \\ \hline
        \texttt{SpMSpV} & $y(i) = A(i,j) * x(j)$  & SuiteSparse \\ \hline
        \texttt{SpMSpM} & $C(i,j) = A(i,k) * B(k,j)$ & Random 5k-5k \\ \hline
        \texttt{TTM (mode 1)} & $C(i,j,l) = A(i,j,k) * B(k,r)$ & Various Real-World Tensors \\ \hline
    \end{tabular}
    \label{tab:computation}
\end{table}

The layouts 
in Table~\ref{tab:layout-defs} can be used in any contraction computation expressed in tensor index notation. Because this work focuses on combining layouts and computation, we 
demonstrate these layouts in the context of the 
computations in Table~\ref{tab:computation}. 
Sparse matrix vector multiply (\texttt{SpMV}) demonstrates sparse-dense co-iteration and provides the baseline performance of the generated code. \texttt{SpMSpV} deals with sparse-sparse co-iteration where only a single vector is co-iterated with each row. \texttt{SpMSpM} showcases the composability of our polyhedral-framework-based methods when multiple sparse layouts are involved, which  
results in complex loop conditions. We also added tensor-times-matrix (\texttt{TTM}) where the tensor is stored in compressed sparse fiber layout~\cite{csf} (\texttt{CSF}).

These sets of experiments were run on a single-socket 
AMD EPYC 7702P CPUs at 2.0GHz. 
This CPU exposes 4 NUMA domains corresponding to the 4 quadrants, each containing 16 cores, and has its own DRAM controller. Experiments measure multi-threaded code bound
to one of the NUMA domains using the \texttt{numactl} utilities to prevent adverse NUMA effects from suboptimal thread placement. 

The generated code is automatically parallelized by inserting the OpenMP pragma, \texttt{\#pragma omp parallel for} \texttt{schedule (dynamic,32)}, at the outermost parallel loop based on static dependence testing.
All code is compiled with GCC 10.2.0, with flags
\texttt{-O3 -ffast-math -march=native}. 

We compared the generated code 
with corresponding kernels from a state-of-the-art tensor algebra compiler --- TACO~\cite{taco}, an optimized binary library --- the Intel Math Kernel Libraries~\cite{mkl} 2021.1.1, a template library --- Eigen~\cite{eigenweb} 3.3.7, and a state-of-the-art sparse linear algebra library --- SuiteSparse:GraphBLAS~\cite{suiteparsegraphblas} (SS:GB) 5.10.1. Parallel implementations from the libraries are used when available. TACO is parallelized using the same OpenMP pragma to eliminate any difference arising from parallelization.

\label{sec:exp-computation}

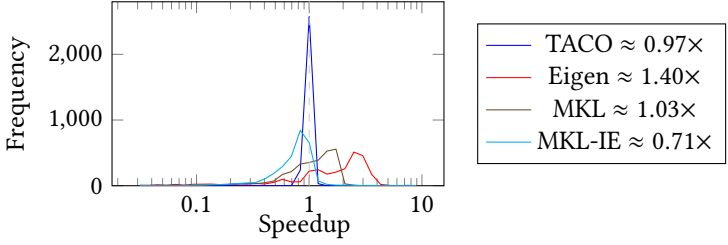
\begin{figure}[t]
\centering
\begin{tikzpicture}
    \begin{axis}[
    ylabel style={align=center},
    ylabel=Frequency, 
      xlabel near ticks,
   ylabel near ticks,
   x label style={at={(axis description cs:0.58,-0.1)},anchor=north},
    xlabel={Speedup}, height=4cm, width=6cm, log ticks with fixed point,
    xmode=log,
    ymin = 0,
    ymax = 2800,
    legend style={at={(1.1, 0.5)}, anchor=west, legend columns=1}]
    \addplot +[mark=none] table [x=speedup, y=taco]{data/spmvdist.dat};
    \addlegendentry{TACO $\approx 0.97\times$};
    \addplot +[mark=none] table [x=speedup, y=eigen]{data/spmvdist.dat};
    \addlegendentry{Eigen $\approx 1.40\times$};
    \addplot +[mark=none] table [x=speedup, y=mkl]{data/spmvdist.dat};
    \addlegendentry{MKL $\approx 1.03\times$};
    \addplot +[color=cyan, mark=none] table [x=speedup, y=mklie]{data/spmvdist.dat};
    \addlegendentry{MKL-IE $\approx 0.71\times$};
    \addplot [black!20, dashed, mark=none] coordinates {(1,0.1) (1, 3000)};
    \end{axis}
\end{tikzpicture}
\caption{Distribution of relative multi-threaded SpMV execution time on SuiteSparse matrices: we are able to generate code of similar quality when sparse-sparse co-iteration is not concerned. MKL-IE is the inspector executor version of MKL. Each bucket is of size $1.2$: bucket $x$ is $[1.2^{x-0.5},1.2^{x+0.5})$; the bucket around $1$ --- no speedup, is $[0.913,1.095)$.}
\label{fig:spmv}
\end{figure}

\subsection{Performance without co-iteration: SpMV}

Fig.~\ref{fig:spmv} provides the performance comparison 
on 2893 of the the real and pattern matrices in the SuiteSparse matrix collection~\cite{suitesparse} in CSR layout for all libraries. Each experiment is run at least two times or until 30 seconds have elapsed. The average speedup 
is reported by geometric mean over the speedup from all random sets.

Considering \texttt{SpMV}, we see comparable performance to TACO and the library implementations
other than MKL-IE, demonstrating that code generation using polyhedral scanning and synthesis achieves efficient code.
MKL-IE achieves higher performance using a runtime inspector that can fine-tune the loop schedules and parallelization strategy of the executor; inspection time is not included in the execution time measurement.

\begin{table}[t]
    \centering
        \caption{We achieved significant speedup (geometric mean) on SpMSpV 
    where sparse-sparse co-iteration is involved. We present results for sequential iteration (\texttt{SeqIter}) or hashmap (\texttt{HashMap}) as find algorithms.  \texttt{Auto} selects the best performance using these two algorithms.}  
    \begin{tabular}{c|c|c|c} \hline
          & \textbf{TACO} & \textbf{Eigen} & \textbf{SS:GB} \\ \hline \hline
        \texttt{SeqIter} & 1.63 & 2.43 & 1.50 \\ \hline
        \texttt{HashMap} & 1.63 & 2.44 & 1.50 \\ \hline
        \texttt{Auto} & 2.72 & 4.06 & 2.50 \\ \hline
    \end{tabular}

    \label{tab:spmv-real}
\end{table}

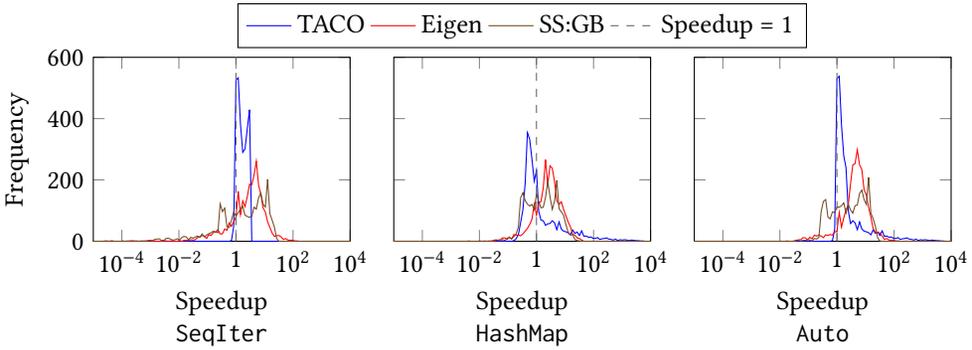
\begin{figure}[t]
\centering
\begin{tikzpicture}
\pgfplotsset{
    height=4cm, width=5cm, 
}
    \begin{axis}[
    name=plot1,
    log ticks with fixed point,
   xmode=log,
    ymin = 0,
    ymax = 600,
    ylabel style={align=center},
    ylabel=Frequency, 
   xmin = 0.00001,
   xmax = 10000,
   log basis x={10},
   xtick = {0.0001,0.01,1,100, 10000},
   xticklabels = {$10^{-4}$, $10^{-2}$, $1$, $10^{2}$, $10^{4}$},
   xlabel style={align=center},
    xlabel={Speedup \\ \texttt{SeqIter}}
    ]
    \addplot +[mark=none] table [x=perfv, y=taco]{data/distseq.dat};
    \addplot +[mark=none] table [x=perfv, y=eigen]{data/distseq.dat};
    \addplot +[mark=none] table [x=perfv, y=graphblas]{data/distseq.dat};
    \addplot [black!60, dashed, mark=none] coordinates {(1,0.1) (1, 3000)};
    \end{axis}
    \begin{axis}[at={(4cm,0)}, anchor=south west, 
    ymin = 0,
    ymax = 600,
    yticklabels={,,},
   xmode=log,log ticks with fixed point,
   xlabel style={align=center},
    xlabel={Speedup \\ \texttt{HashMap}},
   log basis x={10},
   xmin = 0.00001,
   xmax = 10000,
   xtick = {0.0001,0.01,1,100, 10000},
   xticklabels = {$10^{-4}$, $10^{-2}$, $1$, $10^{2}$, $10^{4}$},
    legend style={at={(0.5, 1.05)}, anchor=south, legend columns=4}]
    \addplot +[mark=none] table [x=perfv, y=taco]{data/disthash.dat};
    \addlegendentry{TACO};
    \addplot +[mark=none] table [x=perfv, y=eigen]{data/disthash.dat};
    \addlegendentry{Eigen};
    \addplot +[mark=none] table [x=perfv, y=graphblas]{data/disthash.dat};
    \addlegendentry{SS:GB};
    \addplot [black!60, dashed, mark=none] coordinates {(1,0.1) (1, 3000)};
    \addlegendentry{Speedup = 1}
    \end{axis}
    \begin{axis}[at={(8cm,0)}, anchor=south west, 
    ymin = 0,
    ymax = 600,
   xmode=log,log ticks with fixed point,
    yticklabels={,,},
    xlabel style={align=center},
    xlabel={Speedup \\ \texttt{Auto}}, 
   xmin = 0.00001,
   xmax = 10000,
   log basis x={10},
   xtick = {0.0001,0.01,1,100, 10000},
   xticklabels = {$10^{-4}$, $10^{-2}$, $1$, $10^{2}$, $10^{4}$},
    ]
    \addplot +[mark=none] table [x=perfv, y=taco]{data/dist.dat};
    \addplot +[mark=none] table [x=perfv, y=eigen]{data/dist.dat};
    \addplot +[mark=none] table [x=perfv, y=graphblas]{data/dist.dat};
    \addplot [black!60, dashed, mark=none] coordinates {(1,0.1) (1, 3000)};
    \end{axis}
\end{tikzpicture}
\caption{Distribution of relative speedup of generated code to different libraries: \texttt{Auto} is consistently faster than other libraries. The right side of $1$ indicates the implementation is able to obtain a speedup greater than $1\times$. Frequency denote the number of matrices from SuiteSparse that have relative speedup in the bucket. Each bucket is of size $1.2$: bucket $x$ is $[1.2^{x-0.5},1.2^{x+0.5})$; the bucket around $1$ --- no speedup, is $[0.913,1.095)$.}
\label{fig:spmspvopt}
\end{figure}

\begin{figure}[t!]
\centering
\includegraphics{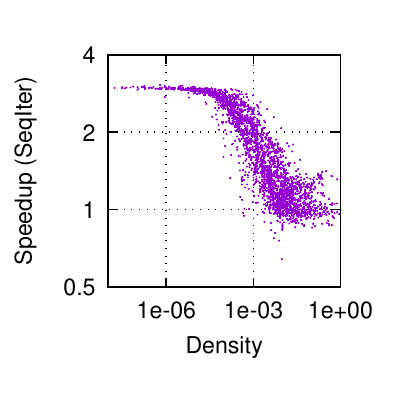}
\includegraphics{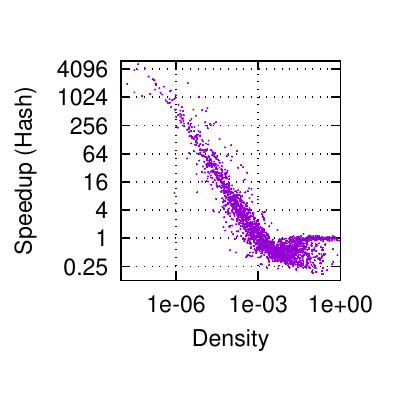}
\includegraphics{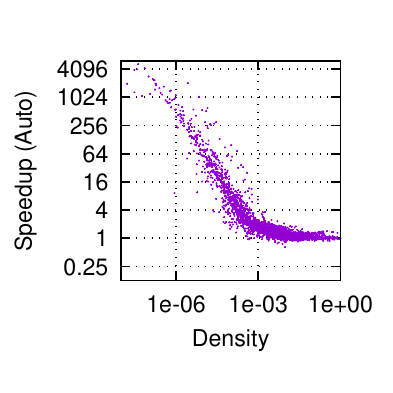}
\vspace{-1em}
\caption{Speedup against TACO under different matrix density. Due to less data movement, \texttt{SeqIter} is almost consistently more efficient than TACO's conjunctive merge. \texttt{HashMap} is of different algorithmic complexity compared to both TACO and \texttt{SeqIter}. By offering the flexibility of generating either algorithms, \texttt{Auto}, our code generation algorithm achieves non-constant speedup against TACO.}
\label{fig:v-taco}
\end{figure}

\subsection{Performance of co-iteration: SpMSpV}

\texttt{SpMSpV} demonstrates the performance of generated code on sparse-sparse co-iteration. We exclude $17$ of the matrices due to either out-of-memory issues caused by SS:GB when performing inspections or time-out (longer than 4 hours) caused by TACO that also affects the sequential iteration to a lesser degree. 
The geometric mean speedup achieved is shown in Table~\ref{tab:spmv-real} with the distribution of speedup shown in Figure~\ref{fig:spmspvopt}.

For the individual find algorithms, we are able to achieve consistent performance improvement over the libraries. \texttt{SeqIter}, are consistently faster than the comparable conjunctive merge algorithm from TACO, where both have an algorithm complexity of $O(rows * x.nnz + A.nnz)$. $nnz$ stands for the number of nonzeros in the respective tensor. This improvement is from reduced data movement, as discussed in Section~\ref{sec:co-iteration-analysis}. Meanwhile, \texttt{HashMap} has an algorithm complexity of $O(A.nnz + x.nnz)$ which is much more efficient when the vector is denser than the matrix. The density is defined as the ratio of nonzeros over the size of the tensor in the logical space. Due to different complexity, \texttt{HashMap} achieves the largest speedup of $5214\times$ compared to TACO on the DIMACS10/europe\_osm matrix, but its speedup is also less consistent. Eigen and SuiteSparse:GraphBlas (SS:GB) libraries use computation methods that have a more stable algorithmic complexity related to the number of nonzeros in the sparse matrix, which is comparable to our generated code with \texttt{HashMap}. Furthermore, due to the flexibility of generating different find algorithms, we are able to achieve even higher speedup by selecting the best performing algorithm at runtime.

As discussed, density is an important metric that affects the relative performance of different find algorithms. Fig.~\ref{fig:v-taco} demonstrate its effects on different algorithms when compared to TACO. \texttt{SeqIter} is able to achieve a bounded speedup when the density of the matrix is low, up to $3.05\times$. Meanwhile, \texttt{HashMap} is less affected by the density of the sparse vector. It can achieve unbounded speedup proportional to the differences in density of the sparse matrix and the sparse vector's density of $0.1$. By selecting the best performing algorithm, we are able to achieve consistent speedup by avoiding the slowdowns from \texttt{HashMap} when the matrix density is high and improving the speedup potential when the matrix density is low.

\begin{table}[t]
    \centering
        \caption{\texttt{SpMSpM} involving two sparse matrices.  We show speedup over geometric means as compared to TACO.  $\cmark$ indicates representations not supported by TACO, including 
    lower triangular and warp sparse matrix storage. 
    For the blocked representation, BCSR, we show TACO results only against BCSR, since TACO must represent BCSR as a fourth-order tensor, requiring changes to the other tensor to express as a fourth-order tensor as well. \xmark~ indicates computations that are not supported by TACO while we support all layout combinations.}
    \begin{tabular}{c|c|c|c|c} \hline
        \textbf{Layout A\textbackslash Layout B} & COO & CSR & DCSR & BCSR* \\ \hline \hline
        COO & 1.21 & 1.00 & 0.98 & \xmark \\ \hline
        CSR & 0.99 & 0.99 & 1.00 & \xmark \\ \hline
        DCSR & 0.97 & 1.00 & 1.00 & \xmark \\ \hline
        BCSR & \xmark & \xmark & \xmark & 1.01 \\ \hline
    \end{tabular}

    \label{tab:spmspm-random}
\end{table}

\subsection{Composing programs with multiple layouts: SpMSpM}
We next consider \texttt{SpMSpM} and analyze performance of co-iteration when the two sparse matrices have different layouts.  As we want to look at layouts beyond CSR, and neither Eigen nor SS:GB support other layouts, this comparison is only with TACO.
We used randomly-generated matrices from TACO~\footnote[3]{TACO retrieved from \url{https://github.com/tensor-compiler/taco}, master@c9bd10d6, committed on March 17, 2021. Tensors generated with \texttt{taco::util::fillTensor(tensor, taco::util::FillMethod::Sparse, 0.1)}} with a fillrate 0.1 in each dimension. We report a geometric mean over 100 sets of random tensors in each experiment, where each set is run twice to avoid a cold start and run at least 8 times and at least 5 seconds.

Table~\ref{tab:spmspm-random} demonstrates that our approach is 
able to compose complex layouts with computation while achieving comparable performance to TACO on computations that TACO supports. Our approach improves upon TACO's performance on COO by avoiding multiple sweeps over the sparse tensors.
By decoupling the iteration over logical and sparse indices, our approach is able to support triangular layout and computations and blocked layouts such as BCSR.  To support BCSR in TACO, due to the requirement of matching dimensions to iterators, the BCSR layout must be expressed as a fourth-order tensor that is incompatible with the other layouts besides BCSR\cite{taco}.

\subsection{Performance of higher order tensors and sharing of sparsity structure}

\texttt{TTM} is commonly used in popular tensor decompositions, such as the Tucker decomposition, for a variety of applications, including (social network, electrical grid) data analytics, numerical simulation, machine learning, recommendation systems, personalized web search, etc.~\cite{Kolda:2009:survey,Cichocki:2016:survey,Anandkumar:2014:survey,Sidiropoulos:2017:survey}. In this experiment we use the mode-1 variant of the computation, where the third dimension of the tensor input is contracted, $C(i,j,l) = A(i,j,k) * B(k,r)$. Note that $r$ is typically much smaller than $k$ in low-rank decompositions, typically $r < 100$. 

\texttt{TTM} represents a case where there is known output sparsity when tensor A and C are stored in compressed sparse fiber format: $A(i,j,:) \neq \emptyset \rightarrow C(i,j,:) \neq \emptyset$. With our layout specification, we can describe A and C using the same auxiliary index arrays and having the same sparsity structure for the leading two dimensions.

\begin{table}[t]
    \caption{TTM (mode-1, $R=16$) performance comparison with TACO. \texttt{NNZ} is the number of nonzeros. TACO is unable to generate parallel code due to it sequentially advance in the positions of the sparse output. }
    \centering
    \begin{tabular}{c|c|c|c|c|c}
        \textbf{Tensor} & \textbf{Collection} & \textbf{NNZ} & \textbf{TACO} & \textbf{Generated}  & \textbf{Generated Parallel}\\ \hline \hline
        \multicolumn{6}{c}{\textit{Social Network Analysis}} \\\hline \hline
        \emph{delicious-3d} & FROSTT & 140M & $2.07s$ & $2.14s$ & $0.44s$ \\\hline
        \emph{flickr-3d} & FROSTT & 113M & $1.12s$ & $1.20s$ & $0.27s$ \\\hline
        \emph{freebase-music} & HaTen2 & 100M & $1.26s$ & $1.30s$ & $0.74s$ \\\hline \hline
        \multicolumn{6}{c}{\textit{Pattern Recognition}} \\\hline \hline
        \emph{vast-2015-mc1} & FROSTT & 26M & $0.31s$ & $0.32s$ & $0.19s$ \\\hline \hline
        \multicolumn{6}{c}{\textit{Natural Language Processing}} \\\hline \hline
        \emph{NELL1} & FROSTT & 144M & $8.38s$ & $9.28s$ & $0.91s$ \\\hline
        \emph{NELL2} & FROSTT & 77M & $0.49s$ & $0.49s$ & $0.04$ \\\hline \hline
        \multicolumn{6}{c}{\textit{Anomaly Detection}} \\\hline \hline
        \emph{1998darpa} & HaTen2 & 28M & $0.77s$ & $0.87s$ & $0.20s$ \\\hline
    \end{tabular}
    \label{tab:ttm}
    \vspace{-1em}
\end{table}

Table~\ref{tab:ttm} shows the performance of our generated code against TACO. The sparse tensors, in the compressed sparse fiber (CSF) layout~\cite{csf}, are taken from real-world applications available in The Formidable Repository of Open Sparse Tensors and Tools (FROSTT)~\cite{frosttdataset} and the HaTen2 dataset~\cite{Jeon:2015:haten2}. TACO and our work require pre-generated sparsity using an assemble phase. The timing of assemble is excluded from the table. The performance of our generated code and TACO is similar. However, we are a little slower by reading one more index array for the memory location of the output variable, whereas TACO uses a counter variable for the location. However, TACO cannot generate a parallel compute code due to sequentially writing to the output sparsity structure. In contrast, we can generate efficient parallelizable implementations from sparsity sharing and are, on a geometric average, $4.27\times$ faster.


\subsection{Code Generation Performance}

Table~\ref{tab:codegentime} demonstrates that code generation time is practical compared to compiling the generated code using GCC for the most complex examples. Because tensor contraction is a single perfectly-nested loop nest, polyhedral scanning and synthesizing finds are very efficient. Even one of the most complex codes that compute
\texttt{SpMSpM} using BCSR and COO layout takes less than 0.1 seconds to generate. 

\begin{table}[t]
    \centering
        \caption{Time comparison of code generation and compiling the generated code with GCC in seconds. Codegen is the time from parsing the user-provided input until a C++ source program is produced, including time spent in ILP and SMT solvers.}
    \begin{tabular}{c|c|c}\hline
        Computation & Codegen & GCC  \\\hline
        SpMV(CSR) & 0.036 & 0.182 \\\hline
        SpMV(LowerTri) & 0.036 & 0.181 \\\hline
        SpMSpM(BCSR$\times$COO) & 0.091 & 0.194 \\\hline
    \end{tabular}
    \label{tab:codegentime}
\end{table}


\section{Related work}

Our work can be considered as
an extension to the Sparse Polyhedral Framework. This work presents a layout specification that can be integrated with the polyhedral framework and a code generation algorithm that combines polyhedral scanning with code synthesis of find algorithms.

\subsection{Layout Specification}

There is little work targeted to general layout specification due to it having strong ties to a specific code generation strategy. Sparse libraries may expect a standard layout for the whole tensor \citep{eigenweb,mkl,2020SciPy-NMeth}. 
Compiler-based approaches can vary the sparse implementation of specific dimensions of the tensor using 
a set of names to identify known formats. \citet{aartbik,aartthesis,sipr} defines the implementation of each index dimension of a tensor using a set of names. This approach is refined with TACO~\citep{taco} by~\citet{tacoformat}.

Unlike these prior sparse tensor compilers, our work describes the spaces of the layouts separated from the space of the computation. This enables working with blocked layouts that have a 2:1 mapping from layout dimensions to coordinate dimensions and experimenting with loop orders involving dimensions from layouts and those from the computation.

\subsection{Code Generation}
Tensor contraction engine~\citep{tensorcontractionengine} is a pioneering work to automatically generate dense tensor contraction computations in quantum chemistry, used extensively in the NWChem software suite~\citep{NWChem}. It can automatically determine the binary contraction order for multiple tensor contractions with minimal operation and memory cost, and define and reuse intermediate contraction results.

\citet{aartbik,aartthesis} described a compiler that can transform dense loops over dense arrays into sparse loops over nonzero elements using a technique called guard encapsulation. It treats the index set of an input tensor as a whole to define guard conditions that either include or exclude the inner computation. The Bernoulli project \citep{kotlyar1997compiling,kotlyar1997sparse,kotlyar1997relational,stodghill1997thesis,mateev2000} generates sparse algebra computations by modeling the iterations as DO-ANY loops and formulates the computation as query expressions. It introduces \emph{external} fields to represent dimensions not part of the index coordinates and index hierarchy for preferred ordering of enumeration within a layout. These previous works on sparse tensor algebra are refined with TACO \citep{taco}, which formalizes the dependence between indices using the iteration graph, defines merge lattices for co-iteration, and uses a set of level format and level functions to describe the computation. \citet{sipr} used Enumerator/Accessors to guide the choice of layouts and code generation for sparse computation.

Other frameworks leverage the inspector-executor pattern to achieve data and computation optimization targeting specific sparsity patterns in the tensor. Sparso~\citep{sparso} enables context-driven optimizations using input matrix properties and matrix reordering. Comet~\citep{comet} implements a tensor contraction dialect in Multi-Level IR compiler (MLIR) infrastructure~\citep{lattner2020mlir}. It uses a similar layout specification as TACO and implements data reordering~\citep{reorder} to improve spatial and temporal locality.

\subsection{Polyhedral Frameworks and Sparse Polyhedral Frameworks}

The polyhedral framework can be employed on sparse arrays that does not employ uninterpreted functions. \citet{piecewise} used trace reconstruction to exploit regular patterns in the sparse matrix. Sublimation~\citep{sublimation} turns irregular accesses into regular accesses by exploiting the injectivity of access functions and expanding the irregular loops to regular loops to cover a larger range of values.
Zhao et al. generate code for non-affine loop bounds using conditions and exits\citep{zhaocc18}. There is also prior work in the polyhedral framework that deals with while loops\cite{Benabderrahmane,collard94}. However, these works are insufficient to express co-iterations, which have not only dynamic loop bounds but also dynamic conditions with multiple index arrays.  

The Sparse Polyhedral Framework uses inspectors to analyze properties of sparse computations at runtime in order to create suitable sparse data structures and transform the original computation into executors that can use the new sparse data structure. \citet{anandpldi} described transformation recipes in the polyhedral framework with make-dense and compact operations to compose layout with other loop transformations. \citet{wavefront} described an approach that leverages runtime dependence analysis and layout transformation to achieve wavefront execution of sparse computation. \citet{mahdiinspector} used index array properties to simplify runtime dependence checking and generate efficient inspectors. The polyhedral framework composes the relations provided to the SMT solver and simplifies the dependences used to derive the inspector.

None of these prior works systematically present a specification of sparse layout in the polyhedral framework, and none of them supported co-iterations.

\subsection{Comparison}

\begin{table}[t]
    \centering
        \caption{Comparing this work with related works in sparse polyhedral framework (SPF) and TACO. Support of disjunctive co-iteration ($vee$) and other pattern in \citet{HentrySparseArray2021} is left for future work (F). *Our work enables implementing a co-iteration with a combination of different algorithms: algorithm selection (AS). }
    \begin{tabular}{c|c|c|c|c|c|c|c}\hline
        \multirow{2}{*}{Related Works} & \multicolumn{3}{c|}{Co-iteration} & Layout & Nonzero &  \multicolumn{2}{c}{Index transformations} \\ \cline{2-4} \cline{7-8}
        & $\wedge$ & $\vee$ & AS* & Description & ordering & Blocking & Subsection \\ \hline
        SPF~\cite{anandpldi,wavefront,mahdiinspector} & \xmark & \xmark & \xmark & \xmark & \cmark & Limited & \cmark  \\\hline
        TACO~\cite{taco,tacoformat,HentrySparseArray2021} & \cmark & \cmark & \xmark & \cmark  & Limited &  Limited  & \xmark \\\hline
        This work & \cmark & F & \cmark & \cmark & \cmark & \cmark & \cmark \\\hline
    \end{tabular}
    \vspace{-1em}
    \label{tab:magic}
\end{table}

Table~\ref{tab:magic} compares this work with both sparse polyhedral frameworks and TACO~\cite{taco}. Compared to prior works of sparse polyhedral frameworks, we are the first to enable sparse layout description and support co-iteration generally. These advances means that no comparison with prior works in SPF is possible. Compared to tensor algebra compilers such as TACO~\cite{taco}, we eliminated many special cases in the compiler design and extended the capability in both supporting layouts and implementing co-iteration. However, in this work, the support for disjunctive merge, which can be represented as sparse loop fusion, is future work.

\section{Conclusion \& Future Work}

The polyhedral framework provides mathematical descriptions of loop nest computations that enable dependence testing, composing code transformation sequences, and generating code.  This paper 
similarly achieves this result for sparse tensor co-iteration by extending the polyhedral framework in two key ways: (1) we employ a relation from a sparse tensor layout to its logical coordinate space and compose this with the logical iteration space to derive the sparse iteration space; (2) we implement co-iteration by iterating the layout of one sparse tensor and looking up the indices the other layout through synthesizing a \emph{find} algorithm.  


This work adds another dimension of interaction in automatic tensor code generation with architecture features like single instruction/multiple data (SIMD) or tensor cores. Prior works on tensor blocking and reordering~\cite{piecewise,reorder} can be orthogonally combined to provide computation speedups using such features. As the find algorithms are synthesized in this work, we can also leverage such hardware features for implementing architecture-specific finds. 
Leveraging vectorization for index comparisons in the find algorithm can provide critical speedups when the indices are sparse.

With hardware architectures' increased diversity in functional units, computation capability, and memory bandwidth, adapting data layout and computation to hardware requirements is crucial to achieving performance portability for sparse tensor computations. By proposing a flexible framework for layout description and code transformation, we have opened up more opportunities for the co-optimization of layout and computation. We believe this work is a critical step in automatically generating architecture-specific variants of data layouts and computation programs.

\begin{acks}                            

This research was supported in part by the Exascale Computing Project (17-SC-20-SC), a joint project of the U.S. Department of Energy's Office of Science and National Nuclear Security Administration and by the National Science Foundation under project CCF-2107556.

We would also like to thank John Jolly and Mahesh Lakshminarasimhan, PhD students at the University of Utah, for their help in collecting experiment results.

\end{acks}

\bibliography{bibfile}


\begin{thebibliography}{83}


\ifx \showCODEN    \undefined \def \showCODEN     #1{\unskip}     \fi
\ifx \showDOI      \undefined \def \showDOI       #1{#1}\fi
\ifx \showISBNx    \undefined \def \showISBNx     #1{\unskip}     \fi
\ifx \showISBNxiii \undefined \def \showISBNxiii  #1{\unskip}     \fi
\ifx \showISSN     \undefined \def \showISSN      #1{\unskip}     \fi
\ifx \showLCCN     \undefined \def \showLCCN      #1{\unskip}     \fi
\ifx \shownote     \undefined \def \shownote      #1{#1}          \fi
\ifx \showarticletitle \undefined \def \showarticletitle #1{#1}   \fi
\ifx \showURL      \undefined \def \showURL       {\relax}        \fi
\providecommand\bibfield[2]{#2}
\providecommand\bibinfo[2]{#2}
\providecommand\natexlab[1]{#1}
\providecommand\showeprint[2][]{arXiv:#2}

\bibitem[Abadi et~al\mbox{.}(2016)]%
        {tensorflow}
\bibfield{author}{\bibinfo{person}{Martin Abadi}, \bibinfo{person}{Paul
  Barham}, \bibinfo{person}{Jianmin Chen}, \bibinfo{person}{Zhifeng Chen},
  \bibinfo{person}{Andy Davis}, \bibinfo{person}{Jeffrey Dean},
  \bibinfo{person}{Matthieu Devin}, \bibinfo{person}{Sanjay Ghemawat},
  \bibinfo{person}{Geoffrey Irving}, \bibinfo{person}{Michael Isard},
  \bibinfo{person}{Manjunath Kudlur}, \bibinfo{person}{Josh Levenberg},
  \bibinfo{person}{Rajat Monga}, \bibinfo{person}{Sherry Moore},
  \bibinfo{person}{Derek~G. Murray}, \bibinfo{person}{Benoit Steiner},
  \bibinfo{person}{Paul Tucker}, \bibinfo{person}{Vijay Vasudevan},
  \bibinfo{person}{Pete Warden}, \bibinfo{person}{Martin Wicke},
  \bibinfo{person}{Yuan Yu}, {and} \bibinfo{person}{Xiaoqiang Zheng}.}
  \bibinfo{year}{2016}\natexlab{}.
\newblock \showarticletitle{TensorFlow: A system for large-scale machine
  learning}. In \bibinfo{booktitle}{\emph{12th USENIX Symposium on Operating
  Systems Design and Implementation (OSDI 16)}}. \bibinfo{pages}{265--283}.
\newblock
\urldef\tempurl%
\url{https://www.usenix.org/system/files/conference/osdi16/osdi16-abadi.pdf}
\showURL{%
\tempurl}


\bibitem[affiliates(2021)]%
        {ptx}
\bibfield{author}{\bibinfo{person}{NVIDIA Corporation~\& affiliates}.}
  \bibinfo{year}{2021}\natexlab{}.
\newblock \bibinfo{title}{Parallel Thread Execution ISA Version 7.5}.
\newblock
  \bibinfo{howpublished}{\url{https://docs.nvidia.com/cuda/parallel-thread-execution/index.html}}.
\newblock


\bibitem[Alur et~al\mbox{.}(2013)]%
        {synthesis}
\bibfield{author}{\bibinfo{person}{Rajeev Alur}, \bibinfo{person}{Rastislav
  Bodik}, \bibinfo{person}{Garvit Juniwal}, \bibinfo{person}{Milo M.~K.
  Martin}, \bibinfo{person}{Mukund Raghothaman}, \bibinfo{person}{Sanjit~A.
  Seshia}, \bibinfo{person}{Rishabh Singh}, \bibinfo{person}{Armando
  Solar-Lezama}, \bibinfo{person}{Emina Torlak}, {and}
  \bibinfo{person}{Abhishek Udupa}.} \bibinfo{year}{2013}\natexlab{}.
\newblock \showarticletitle{Syntax-guided synthesis}. In
  \bibinfo{booktitle}{\emph{2013 Formal Methods in Computer-Aided Design}}.
  \bibinfo{pages}{1--8}.
\newblock
\urldef\tempurl%
\url{https://doi.org/10.1109/FMCAD.2013.6679385}
\showDOI{\tempurl}


\bibitem[Anandkumar et~al\mbox{.}(2014)]%
        {Anandkumar:2014:survey}
\bibfield{author}{\bibinfo{person}{Animashree Anandkumar},
  \bibinfo{person}{Rong Ge}, \bibinfo{person}{Daniel Hsu},
  \bibinfo{person}{Sham~M. Kakade}, {and} \bibinfo{person}{Matus Telgarsky}.}
  \bibinfo{year}{2014}\natexlab{}.
\newblock \showarticletitle{Tensor Decompositions for Learning Latent Variable
  Models}.
\newblock \bibinfo{journal}{\emph{J. Mach. Learn. Res.}} \bibinfo{volume}{15},
  \bibinfo{number}{1} (\bibinfo{date}{Jan.} \bibinfo{year}{2014}),
  \bibinfo{pages}{2773--2832}.
\newblock
\showISSN{1532-4435}


\bibitem[Ancourt and Irigoin(1991)]%
        {scanning}
\bibfield{author}{\bibinfo{person}{Corinne Ancourt} {and}
  \bibinfo{person}{Fran\c{c}ois Irigoin}.} \bibinfo{year}{1991}\natexlab{}.
\newblock \showarticletitle{Scanning Polyhedra with DO Loops}. In
  \bibinfo{booktitle}{\emph{Proceedings of the Third ACM SIGPLAN Symposium on
  Principles and Practice of Parallel Programming}} (Williamsburg, Virginia,
  USA) \emph{(\bibinfo{series}{PPOPP '91})}. \bibinfo{publisher}{Association
  for Computing Machinery}, \bibinfo{address}{New York, NY, USA},
  \bibinfo{pages}{39–50}.
\newblock
\showISBNx{0897913906}
\urldef\tempurl%
\url{https://doi.org/10.1145/109625.109631}
\showDOI{\tempurl}


\bibitem[Auer et~al\mbox{.}(2006)]%
        {tensorcontractionengine}
\bibfield{author}{\bibinfo{person}{Alexander~A. Auer}, \bibinfo{person}{Gerald
  Baumgartner}, \bibinfo{person}{David~E. Bernholdt}, \bibinfo{person}{Alina
  Bibireata}, \bibinfo{person}{Venkatesh Choppella}, \bibinfo{person}{Daniel
  Cociorva}, \bibinfo{person}{Xiaoyang Gao}, \bibinfo{person}{Robert Harrison},
  \bibinfo{person}{Sriram Krishnamoorthy}, \bibinfo{person}{Sandhya Krishnan},
  \bibinfo{person}{Chi-Chung Lam}, \bibinfo{person}{Qingda Lu},
  \bibinfo{person}{Marcel Nooijen}, \bibinfo{person}{Russell Pitzer},
  \bibinfo{person}{J. Ramanujam}, \bibinfo{person}{P. Sadayappan}, {and}
  \bibinfo{person}{Alexander Sibiryakov}.} \bibinfo{year}{2006}\natexlab{}.
\newblock \showarticletitle{Automatic code generation for many-body electronic
  structure methods: the tensor contraction engine}.
\newblock \bibinfo{journal}{\emph{Molecular Physics}} \bibinfo{volume}{104},
  \bibinfo{number}{2} (\bibinfo{year}{2006}), \bibinfo{pages}{211--228}.
\newblock
\urldef\tempurl%
\url{https://doi.org/10.1080/00268970500275780}
\showDOI{\tempurl}
\showeprint{https://doi.org/10.1080/00268970500275780}


\bibitem[Augustine et~al\mbox{.}(2019)]%
        {piecewise}
\bibfield{author}{\bibinfo{person}{Travis Augustine},
  \bibinfo{person}{Janarthanan Sarma}, \bibinfo{person}{Louis-No\"{e}l
  Pouchet}, {and} \bibinfo{person}{Gabriel Rodr\'{\i}guez}.}
  \bibinfo{year}{2019}\natexlab{}.
\newblock \showarticletitle{Generating Piecewise-Regular Code from Irregular
  Structures}. In \bibinfo{booktitle}{\emph{Proceedings of the 40th ACM SIGPLAN
  Conference on Programming Language Design and Implementation}} (Phoenix, AZ,
  USA) \emph{(\bibinfo{series}{PLDI 2019})}. \bibinfo{publisher}{Association
  for Computing Machinery}, \bibinfo{address}{New York, NY, USA},
  \bibinfo{pages}{625–639}.
\newblock
\showISBNx{9781450367127}
\urldef\tempurl%
\url{https://doi.org/10.1145/3314221.3314615}
\showDOI{\tempurl}


\bibitem[Bader et~al\mbox{.}(2008)]%
        {Bader2008}
\bibfield{author}{\bibinfo{person}{Brett~W. Bader}, \bibinfo{person}{Michael~W.
  Berry}, {and} \bibinfo{person}{Murray Browne}.}
  \bibinfo{year}{2008}\natexlab{}.
\newblock \bibinfo{booktitle}{\emph{Discussion Tracking in Enron Email Using
  PARAFAC}}.
\newblock \bibinfo{publisher}{Springer London}, \bibinfo{address}{London},
  \bibinfo{pages}{147--163}.
\newblock
\showISBNx{978-1-84800-046-9}
\urldef\tempurl%
\url{https://doi.org/10.1007/978-1-84800-046-9_8}
\showDOI{\tempurl}


\bibitem[Baghdadi et~al\mbox{.}(2019)]%
        {tiramisu}
\bibfield{author}{\bibinfo{person}{Riyadh Baghdadi}, \bibinfo{person}{Jessica
  Ray}, \bibinfo{person}{Malek~Ben Romdhane}, \bibinfo{person}{Emanuele
  Del~Sozzo}, \bibinfo{person}{Abdurrahman Akkas}, \bibinfo{person}{Yunming
  Zhang}, \bibinfo{person}{Patricia Suriana}, \bibinfo{person}{Shoaib Kamil},
  {and} \bibinfo{person}{Saman Amarasinghe}.} \bibinfo{year}{2019}\natexlab{}.
\newblock \showarticletitle{Tiramisu: A Polyhedral Compiler for Expressing Fast
  and Portable Code}. In \bibinfo{booktitle}{\emph{Proceedings of the 2019
  IEEE/ACM International Symposium on Code Generation and Optimization}}
  (Washington, DC, USA) \emph{(\bibinfo{series}{CGO 2019})}.
  \bibinfo{publisher}{IEEE Press}, \bibinfo{pages}{193–205}.
\newblock
\showISBNx{9781728114361}


\bibitem[Benabderrahmane et~al\mbox{.}(2010)]%
        {Benabderrahmane}
\bibfield{author}{\bibinfo{person}{Mohamed-Walid Benabderrahmane},
  \bibinfo{person}{Louis-No{\"e}l Pouchet}, \bibinfo{person}{Albert Cohen},
  {and} \bibinfo{person}{C{\'e}dric Bastoul}.} \bibinfo{year}{2010}\natexlab{}.
\newblock \showarticletitle{The Polyhedral Model Is More Widely Applicable Than
  You Think}. In \bibinfo{booktitle}{\emph{Compiler Construction}},
  \bibfield{editor}{\bibinfo{person}{Rajiv Gupta}} (Ed.).
  \bibinfo{publisher}{Springer Berlin Heidelberg}, \bibinfo{address}{Berlin,
  Heidelberg}, \bibinfo{pages}{283--303}.
\newblock
\showISBNx{978-3-642-11970-5}


\bibitem[Bik(1996)]%
        {aartthesis}
\bibfield{author}{\bibinfo{person}{Aart J.~C. Bik}.}
  \bibinfo{year}{1996}\natexlab{}.
\newblock \emph{\bibinfo{title}{Compiler Support for Sparse Matrix
  Computations}}.
\newblock {PhD} dissertation. \bibinfo{school}{Leiden University}.
\newblock


\bibitem[Bik and Wijshoff(1993)]%
        {aartbik}
\bibfield{author}{\bibinfo{person}{Aart J.~C. Bik} {and} \bibinfo{person}{Harry
  A.~G. Wijshoff}.} \bibinfo{year}{1993}\natexlab{}.
\newblock \showarticletitle{Compilation Techniques for Sparse Matrix
  Computations}. In \bibinfo{booktitle}{\emph{Proceedings of the 7th
  International Conference on Supercomputing}} (Tokyo, Japan)
  \emph{(\bibinfo{series}{ICS '93})}. \bibinfo{publisher}{Association for
  Computing Machinery}, \bibinfo{address}{New York, NY, USA},
  \bibinfo{pages}{416–424}.
\newblock
\showISBNx{089791600X}
\urldef\tempurl%
\url{https://doi.org/10.1145/165939.166023}
\showDOI{\tempurl}


\bibitem[Bondhugula et~al\mbox{.}(2008)]%
        {pluto}
\bibfield{author}{\bibinfo{person}{Uday Bondhugula}, \bibinfo{person}{Albert
  Hartono}, \bibinfo{person}{J. Ramanujam}, {and} \bibinfo{person}{P.
  Sadayappan}.} \bibinfo{year}{2008}\natexlab{}.
\newblock \showarticletitle{A Practical Automatic Polyhedral Parallelizer and
  Locality Optimizer}. In \bibinfo{booktitle}{\emph{Proceedings of the 29th ACM
  SIGPLAN Conference on Programming Language Design and Implementation}}
  (Tucson, AZ, USA) \emph{(\bibinfo{series}{PLDI '08})}.
  \bibinfo{publisher}{Association for Computing Machinery},
  \bibinfo{address}{New York, NY, USA}, \bibinfo{pages}{101–113}.
\newblock
\showISBNx{9781595938602}
\urldef\tempurl%
\url{https://doi.org/10.1145/1375581.1375595}
\showDOI{\tempurl}


\bibitem[Bradley et~al\mbox{.}(2006)]%
        {whats}
\bibfield{author}{\bibinfo{person}{Aaron~R. Bradley}, \bibinfo{person}{Zohar
  Manna}, {and} \bibinfo{person}{Henny~B. Sipma}.}
  \bibinfo{year}{2006}\natexlab{}.
\newblock \showarticletitle{What's Decidable About Arrays?}. In
  \bibinfo{booktitle}{\emph{Verification, Model Checking, and Abstract
  Interpretation}}, \bibfield{editor}{\bibinfo{person}{E.~Allen Emerson} {and}
  \bibinfo{person}{Kedar~S. Namjoshi}} (Eds.). \bibinfo{publisher}{Springer
  Berlin Heidelberg}, \bibinfo{address}{Berlin, Heidelberg},
  \bibinfo{pages}{427--442}.
\newblock
\showISBNx{978-3-540-31622-0}


\bibitem[Buluc and Gilbert(2008)]%
        {buluc2008}
\bibfield{author}{\bibinfo{person}{Aydin Buluc} {and} \bibinfo{person}{John~R.
  Gilbert}.} \bibinfo{year}{2008}\natexlab{}.
\newblock \showarticletitle{On the representation and multiplication of
  hypersparse matrices}. In \bibinfo{booktitle}{\emph{2008 IEEE International
  Symposium on Parallel and Distributed Processing}}. \bibinfo{pages}{1--11}.
\newblock
\urldef\tempurl%
\url{https://doi.org/10.1109/IPDPS.2008.4536313}
\showDOI{\tempurl}


\bibitem[Canny and Zhao(2013)]%
        {10.1145/2487575.2487677}
\bibfield{author}{\bibinfo{person}{John Canny} {and} \bibinfo{person}{Huasha
  Zhao}.} \bibinfo{year}{2013}\natexlab{}.
\newblock \showarticletitle{Big Data Analytics with Small Footprint: Squaring
  the Cloud}. In \bibinfo{booktitle}{\emph{Proceedings of the 19th ACM SIGKDD
  International Conference on Knowledge Discovery and Data Mining}} (Chicago,
  Illinois, USA) \emph{(\bibinfo{series}{KDD '13})}.
  \bibinfo{publisher}{Association for Computing Machinery},
  \bibinfo{address}{New York, NY, USA}, \bibinfo{pages}{95–103}.
\newblock
\showISBNx{9781450321747}
\urldef\tempurl%
\url{https://doi.org/10.1145/2487575.2487677}
\showDOI{\tempurl}


\bibitem[Chen(2012)]%
        {Chenpldi2012}
\bibfield{author}{\bibinfo{person}{Chun Chen}.}
  \bibinfo{year}{2012}\natexlab{}.
\newblock \showarticletitle{Polyhedra Scanning Revisited}. In
  \bibinfo{booktitle}{\emph{Proceedings of the 33rd ACM SIGPLAN Conference on
  Programming Language Design and Implementation}} (Beijing, China)
  \emph{(\bibinfo{series}{PLDI '12})}. \bibinfo{publisher}{Association for
  Computing Machinery}, \bibinfo{address}{New York, NY, USA},
  \bibinfo{pages}{499–508}.
\newblock
\showISBNx{9781450312059}
\urldef\tempurl%
\url{https://doi.org/10.1145/2254064.2254123}
\showDOI{\tempurl}


\bibitem[Chen et~al\mbox{.}(2008)]%
        {chill}
\bibfield{author}{\bibinfo{person}{Chun Chen}, \bibinfo{person}{Jacqueline
  Chame}, {and} \bibinfo{person}{Mary Hall}.} \bibinfo{year}{2008}\natexlab{}.
\newblock \bibinfo{booktitle}{\emph{CHiLL: A framework for composing high-level
  loop transformations}}.
\newblock \bibinfo{type}{{T}echnical {R}eport}.
  \bibinfo{institution}{Citeseer}.
\newblock


\bibitem[Chen et~al\mbox{.}(2018)]%
        {tvm}
\bibfield{author}{\bibinfo{person}{Tianqi Chen}, \bibinfo{person}{Thierry
  Moreau}, \bibinfo{person}{Ziheng Jiang}, \bibinfo{person}{Lianmin Zheng},
  \bibinfo{person}{Eddie Yan}, \bibinfo{person}{Meghan Cowan},
  \bibinfo{person}{Haichen Shen}, \bibinfo{person}{Leyuan Wang},
  \bibinfo{person}{Yuwei Hu}, \bibinfo{person}{Luis Ceze},
  \bibinfo{person}{Carlos Guestrin}, {and} \bibinfo{person}{Arvind
  Krishnamurthy}.} \bibinfo{year}{2018}\natexlab{}.
\newblock \showarticletitle{TVM: An Automated End-to-End Optimizing Compiler
  for Deep Learning}. In \bibinfo{booktitle}{\emph{Proceedings of the 13th
  USENIX Conference on Operating Systems Design and Implementation}} (Carlsbad,
  CA, USA) \emph{(\bibinfo{series}{OSDI'18})}. \bibinfo{publisher}{USENIX
  Association}, \bibinfo{address}{USA}, \bibinfo{pages}{579–594}.
\newblock
\showISBNx{9781931971478}


\bibitem[Chou et~al\mbox{.}(2018)]%
        {tacoformat}
\bibfield{author}{\bibinfo{person}{Stephen Chou}, \bibinfo{person}{Fredrik
  Kjolstad}, {and} \bibinfo{person}{Saman Amarasinghe}.}
  \bibinfo{year}{2018}\natexlab{}.
\newblock \showarticletitle{Format Abstraction for Sparse Tensor Algebra
  Compilers}.
\newblock \bibinfo{journal}{\emph{Proc. ACM Program. Lang.}}
  \bibinfo{volume}{2}, \bibinfo{number}{OOPSLA}, Article
  \bibinfo{articleno}{123} (\bibinfo{date}{Oct.} \bibinfo{year}{2018}),
  \bibinfo{numpages}{30}~pages.
\newblock
\urldef\tempurl%
\url{https://doi.org/10.1145/3276493}
\showDOI{\tempurl}


\bibitem[{Cichocki} et~al\mbox{.}(2016)]%
        {Cichocki:2016:survey}
\bibfield{author}{\bibinfo{person}{A. {Cichocki}}, \bibinfo{person}{N. {Lee}},
  \bibinfo{person}{I.~V. {Oseledets}}, \bibinfo{person}{A. {Phan}},
  \bibinfo{person}{Q. {Zhao}}, {and} \bibinfo{person}{D. {Mandic}}.}
  \bibinfo{year}{2016}\natexlab{}.
\newblock \showarticletitle{Low-Rank Tensor Networks for Dimensionality
  Reduction and Large-Scale Optimization Problems: Perspectives and Challenges
  PART 1}.
\newblock \bibinfo{journal}{\emph{ArXiv e-prints}} (\bibinfo{date}{Sept.}
  \bibinfo{year}{2016}).
\newblock
\showeprint[arxiv]{1609.00893}~[cs.NA]


\bibitem[Collard(1994)]%
        {collard94}
\bibfield{author}{\bibinfo{person}{J.-F. Collard}.}
  \bibinfo{year}{1994}\natexlab{}.
\newblock \showarticletitle{Space-time transformation of while-loops using
  speculative execution}. In \bibinfo{booktitle}{\emph{Proceedings of IEEE
  Scalable High Performance Computing Conference}}. \bibinfo{pages}{429--436}.
\newblock
\urldef\tempurl%
\url{https://doi.org/10.1109/SHPCC.1994.296675}
\showDOI{\tempurl}


\bibitem[Davis(2019)]%
        {suiteparsegraphblas}
\bibfield{author}{\bibinfo{person}{Timothy~A. Davis}.}
  \bibinfo{year}{2019}\natexlab{}.
\newblock \showarticletitle{Algorithm 1000: SuiteSparse:GraphBLAS: Graph
  Algorithms in the Language of Sparse Linear Algebra}.
\newblock \bibinfo{journal}{\emph{ACM Trans. Math. Softw.}}
  \bibinfo{volume}{45}, \bibinfo{number}{4}, Article \bibinfo{articleno}{44}
  (\bibinfo{date}{Dec.} \bibinfo{year}{2019}), \bibinfo{numpages}{25}~pages.
\newblock
\showISSN{0098-3500}
\urldef\tempurl%
\url{https://doi.org/10.1145/3322125}
\showDOI{\tempurl}


\bibitem[Davis and Hu(2011)]%
        {suitesparse}
\bibfield{author}{\bibinfo{person}{Timothy~A. Davis} {and}
  \bibinfo{person}{Yifan Hu}.} \bibinfo{year}{2011}\natexlab{}.
\newblock \showarticletitle{The University of Florida Sparse Matrix
  Collection}.
\newblock \bibinfo{journal}{\emph{ACM Trans. Math. Softw.}}
  \bibinfo{volume}{38}, \bibinfo{number}{1}, Article \bibinfo{articleno}{1}
  (\bibinfo{date}{Dec.} \bibinfo{year}{2011}), \bibinfo{numpages}{25}~pages.
\newblock
\showISSN{0098-3500}
\urldef\tempurl%
\url{https://doi.org/10.1145/2049662.2049663}
\showDOI{\tempurl}


\bibitem[De~Moura and Bj\o{}rner(2008)]%
        {z3}
\bibfield{author}{\bibinfo{person}{Leonardo De~Moura} {and}
  \bibinfo{person}{Nikolaj Bj\o{}rner}.} \bibinfo{year}{2008}\natexlab{}.
\newblock \showarticletitle{Z3: An Efficient SMT Solver}. In
  \bibinfo{booktitle}{\emph{Proceedings of the Theory and Practice of Software,
  14th International Conference on Tools and Algorithms for the Construction
  and Analysis of Systems}} (Budapest, Hungary)
  \emph{(\bibinfo{series}{TACAS'08/ETAPS'08})}.
  \bibinfo{publisher}{Springer-Verlag}, \bibinfo{address}{Berlin, Heidelberg},
  \bibinfo{pages}{337–340}.
\newblock
\showISBNx{3540787992}


\bibitem[Einstein(1916)]%
        {einstein}
\bibfield{author}{\bibinfo{person}{A. Einstein}.}
  \bibinfo{year}{1916}\natexlab{}.
\newblock \showarticletitle{Die Grundlage der allgemeinen
  Relativitätstheorie}.
\newblock \bibinfo{journal}{\emph{Annalen der Physik}} \bibinfo{volume}{354},
  \bibinfo{number}{7} (\bibinfo{year}{1916}), \bibinfo{pages}{769--822}.
\newblock
\urldef\tempurl%
\url{https://doi.org/10.1002/andp.19163540702}
\showDOI{\tempurl}
\showeprint{https://onlinelibrary.wiley.com/doi/pdf/10.1002/andp.19163540702}


\bibitem[Feautrier(1991)]%
        {feautrier1991dataflow}
\bibfield{author}{\bibinfo{person}{Paul Feautrier}.}
  \bibinfo{year}{1991}\natexlab{}.
\newblock \showarticletitle{Dataflow analysis of array and scalar references}.
\newblock \bibinfo{journal}{\emph{International Journal of Parallel
  Programming}} \bibinfo{volume}{20}, \bibinfo{number}{1}
  (\bibinfo{year}{1991}), \bibinfo{pages}{23--53}.
\newblock


\bibitem[Feautrier(1992a)]%
        {feautrier1992some1}
\bibfield{author}{\bibinfo{person}{Paul Feautrier}.}
  \bibinfo{year}{1992}\natexlab{a}.
\newblock \showarticletitle{Some efficient solutions to the affine scheduling
  problem. I. One-dimensional time}.
\newblock \bibinfo{journal}{\emph{International journal of parallel
  programming}} \bibinfo{volume}{21}, \bibinfo{number}{5}
  (\bibinfo{year}{1992}), \bibinfo{pages}{313--347}.
\newblock


\bibitem[Feautrier(1992b)]%
        {feautrier1992some2}
\bibfield{author}{\bibinfo{person}{Paul Feautrier}.}
  \bibinfo{year}{1992}\natexlab{b}.
\newblock \showarticletitle{Some efficient solutions to the affine scheduling
  problem. II. Multidimensional time}.
\newblock \bibinfo{journal}{\emph{International journal of parallel
  programming}} \bibinfo{volume}{21}, \bibinfo{number}{6}
  (\bibinfo{year}{1992}), \bibinfo{pages}{389--420}.
\newblock


\bibitem[Girbal et~al\mbox{.}(2006)]%
        {girbal06}
\bibfield{author}{\bibinfo{person}{Sylvain Girbal}, \bibinfo{person}{Nicolas
  Vasilache}, \bibinfo{person}{Cédric Bastoul}, \bibinfo{person}{Albert
  Cohen}, \bibinfo{person}{David Parello}, \bibinfo{person}{Marc Sigler}, {and}
  \bibinfo{person}{Olivier Temam}.} \bibinfo{year}{2006}\natexlab{}.
\newblock \showarticletitle{Semi-Automatic Composition of Loop Transformations
  for Deep Parallelism and Memory Hierarchies}.
\newblock \bibinfo{journal}{\emph{International Journal of Parallel
  Programming}}  \bibinfo{volume}{34} (\bibinfo{date}{06}
  \bibinfo{year}{2006}), \bibinfo{pages}{261--317}.
\newblock
\urldef\tempurl%
\url{https://doi.org/10.1007/s10766-006-0012-3}
\showDOI{\tempurl}


\bibitem[Griebl et~al\mbox{.}(1998)]%
        {griebl98}
\bibfield{author}{\bibinfo{person}{M. Griebl}, \bibinfo{person}{C. Lengauer},
  {and} \bibinfo{person}{S. Wetzel}.} \bibinfo{year}{1998}\natexlab{}.
\newblock \showarticletitle{Code generation in the polytope model}. In
  \bibinfo{booktitle}{\emph{Proceedings. 1998 International Conference on
  Parallel Architectures and Compilation Techniques (Cat. No.98EX192)}}.
  \bibinfo{pages}{106--111}.
\newblock
\urldef\tempurl%
\url{https://doi.org/10.1109/PACT.1998.727179}
\showDOI{\tempurl}


\bibitem[Grosser et~al\mbox{.}(2015)]%
        {tobias2015}
\bibfield{author}{\bibinfo{person}{Tobias Grosser}, \bibinfo{person}{Sven
  Verdoolaege}, {and} \bibinfo{person}{Albert Cohen}.}
  \bibinfo{year}{2015}\natexlab{}.
\newblock \showarticletitle{Polyhedral AST Generation Is More Than Scanning
  Polyhedra}.
\newblock \bibinfo{journal}{\emph{ACM Trans. Program. Lang. Syst.}}
  \bibinfo{volume}{37}, \bibinfo{number}{4}, Article \bibinfo{articleno}{12}
  (\bibinfo{date}{July} \bibinfo{year}{2015}), \bibinfo{numpages}{50}~pages.
\newblock
\showISSN{0164-0925}
\urldef\tempurl%
\url{https://doi.org/10.1145/2743016}
\showDOI{\tempurl}


\bibitem[Guennebaud et~al\mbox{.}(2010)]%
        {eigenweb}
\bibfield{author}{\bibinfo{person}{Ga\"{e}l Guennebaud},
  \bibinfo{person}{Beno\^{i}t Jacob}, {et~al\mbox{.}}}
  \bibinfo{year}{2010}\natexlab{}.
\newblock \bibinfo{title}{Eigen v3}.
\newblock \bibinfo{howpublished}{http://eigen.tuxfamily.org}.
\newblock


\bibitem[Hentry et~al\mbox{.}(2021)]%
        {HentrySparseArray2021}
\bibfield{author}{\bibinfo{person}{Rawn Hentry}, \bibinfo{person}{Olivia Hsu},
  \bibinfo{person}{Rohan Yadav}, \bibinfo{person}{Stephen Chou},
  \bibinfo{person}{Kunle Olukotun}, \bibinfo{person}{Saman Amarasinghe}, {and}
  \bibinfo{person}{Fredrik Kjolstad}.} \bibinfo{year}{2021}\natexlab{}.
\newblock \showarticletitle{Compilation of Sparse Array Programming Models}.
\newblock \bibinfo{journal}{\emph{Proc. ACM Program. Lang}}
  \bibinfo{volume}{5}, \bibinfo{number}{OOPSLA} (\bibinfo{date}{October}
  \bibinfo{year}{2021}).
\newblock


\bibitem[Im and Yelick(2001)]%
        {im2001}
\bibfield{author}{\bibinfo{person}{Eun-Jin Im} {and}
  \bibinfo{person}{Katherine~A. Yelick}.} \bibinfo{year}{2001}\natexlab{}.
\newblock \showarticletitle{Optimizing Sparse Matrix Computations for Register
  Reuse in SPARSITY}. In \bibinfo{booktitle}{\emph{Proceedings of the
  International Conference on Computational Sciences-Part I}}
  \emph{(\bibinfo{series}{ICCS '01})}. \bibinfo{publisher}{Springer-Verlag},
  \bibinfo{address}{Berlin, Heidelberg}, \bibinfo{pages}{127–136}.
\newblock
\showISBNx{3540422323}


\bibitem[{Intel Corporation}(2022)]%
        {mkl}
\bibfield{author}{\bibinfo{person}{{Intel Corporation}}.}
  \bibinfo{year}{2022}\natexlab{}.
\newblock \bibinfo{title}{"Intel oneAPI Math Kernel Library," Accessed Apr. 5,
  2022}.
\newblock
\newblock
\urldef\tempurl%
\url{https://www.intel.com/content/www/us/en/developer/tools/oneapi/onemkl.html}
\showURL{%
\tempurl}


\bibitem[Irigoin and Triolet(1988)]%
        {supernode}
\bibfield{author}{\bibinfo{person}{F. Irigoin} {and} \bibinfo{person}{R.
  Triolet}.} \bibinfo{year}{1988}\natexlab{}.
\newblock \showarticletitle{Supernode Partitioning}. In
  \bibinfo{booktitle}{\emph{Proceedings of the 15th ACM SIGPLAN-SIGACT
  Symposium on Principles of Programming Languages}} (San Diego, California,
  USA) \emph{(\bibinfo{series}{POPL '88})}. \bibinfo{publisher}{Association for
  Computing Machinery}, \bibinfo{address}{New York, NY, USA},
  \bibinfo{pages}{319–329}.
\newblock
\showISBNx{0897912527}
\urldef\tempurl%
\url{https://doi.org/10.1145/73560.73588}
\showDOI{\tempurl}


\bibitem[Jeon et~al\mbox{.}(2015)]%
        {Jeon:2015:haten2}
\bibfield{author}{\bibinfo{person}{Inah Jeon}, \bibinfo{person}{Evangelos~E.
  Papalexakis}, \bibinfo{person}{U Kang}, {and} \bibinfo{person}{Christos
  Faloutsos}.} \bibinfo{year}{2015}\natexlab{}.
\newblock \showarticletitle{{HaTen2}: Billion-scale Tensor Decompositions}. In
  \bibinfo{booktitle}{\emph{IEEE International Conference on Data Engineering
  (ICDE)}}.
\newblock


\bibitem[Kelly(1998)]%
        {kelly_optimization_1998}
\bibfield{author}{\bibinfo{person}{Wayne Kelly}.}
  \bibinfo{year}{1998}\natexlab{}.
\newblock \emph{\bibinfo{title}{Optimization within a Unified Transformation
  Framework}}.
\newblock {PhD} dissertation. \bibinfo{school}{University of Maryland}.
\newblock
\urldef\tempurl%
\url{https://drum.lib.umd.edu/handle/1903/865}
\showURL{%
\tempurl}
\newblock
\shownote{Accepted: 2004-05-31T22:43:07Z}.


\bibitem[Khan et~al\mbox{.}(2013)]%
        {cudachill}
\bibfield{author}{\bibinfo{person}{Malik Khan}, \bibinfo{person}{Protonu Basu},
  \bibinfo{person}{Gabe Rudy}, \bibinfo{person}{Mary Hall},
  \bibinfo{person}{Chun Chen}, {and} \bibinfo{person}{Jacqueline Chame}.}
  \bibinfo{year}{2013}\natexlab{}.
\newblock \showarticletitle{A Script-Based Autotuning Compiler System to
  Generate High-Performance CUDA Code}.
\newblock \bibinfo{journal}{\emph{ACM Trans. Archit. Code Optim.}}
  \bibinfo{volume}{9}, \bibinfo{number}{4}, Article \bibinfo{articleno}{31}
  (\bibinfo{date}{Jan.} \bibinfo{year}{2013}), \bibinfo{numpages}{25}~pages.
\newblock
\showISSN{1544-3566}
\urldef\tempurl%
\url{https://doi.org/10.1145/2400682.2400690}
\showDOI{\tempurl}


\bibitem[Kjolstad et~al\mbox{.}(2017)]%
        {taco}
\bibfield{author}{\bibinfo{person}{Fredrik Kjolstad}, \bibinfo{person}{Shoaib
  Kamil}, \bibinfo{person}{Stephen Chou}, \bibinfo{person}{David Lugato}, {and}
  \bibinfo{person}{Saman Amarasinghe}.} \bibinfo{year}{2017}\natexlab{}.
\newblock \showarticletitle{The Tensor Algebra Compiler}.
\newblock \bibinfo{journal}{\emph{Proc. ACM Program. Lang.}}
  \bibinfo{volume}{1}, \bibinfo{number}{OOPSLA}, Article
  \bibinfo{articleno}{77} (\bibinfo{date}{Oct.} \bibinfo{year}{2017}),
  \bibinfo{numpages}{29}~pages.
\newblock
\urldef\tempurl%
\url{https://doi.org/10.1145/3133901}
\showDOI{\tempurl}


\bibitem[Kolda and Bader(2009)]%
        {Kolda:2009:survey}
\bibfield{author}{\bibinfo{person}{T. Kolda} {and} \bibinfo{person}{B. Bader}.}
  \bibinfo{year}{2009}\natexlab{}.
\newblock \showarticletitle{Tensor Decompositions and Applications}.
\newblock \bibinfo{journal}{\emph{SIAM Rev.}} \bibinfo{volume}{51},
  \bibinfo{number}{3} (\bibinfo{year}{2009}), \bibinfo{pages}{455--500}.
\newblock
\urldef\tempurl%
\url{https://doi.org/10.1137/07070111X}
\showDOI{\tempurl}


\bibitem[Kotlyar and Pingali(1997)]%
        {kotlyar1997sparse}
\bibfield{author}{\bibinfo{person}{Vladimir Kotlyar} {and}
  \bibinfo{person}{Keshav Pingali}.} \bibinfo{year}{1997}\natexlab{}.
\newblock \showarticletitle{Sparse code generation for imperfectly nested loops
  with dependences}. In \bibinfo{booktitle}{\emph{Proceedings of the 11th
  international conference on Supercomputing}}. \bibinfo{pages}{188--195}.
\newblock


\bibitem[Kotlyar et~al\mbox{.}(1997a)]%
        {kotlyar1997compiling}
\bibfield{author}{\bibinfo{person}{Vladimir Kotlyar}, \bibinfo{person}{Keshav
  Pingali}, {and} \bibinfo{person}{Paul Stodghill}.}
  \bibinfo{year}{1997}\natexlab{a}.
\newblock \bibinfo{booktitle}{\emph{Compiling parallel sparse code for
  user-defined data structures}}.
\newblock \bibinfo{type}{{T}echnical {R}eport}. \bibinfo{institution}{Cornell
  University}.
\newblock


\bibitem[Kotlyar et~al\mbox{.}(1997b)]%
        {kotlyar1997relational}
\bibfield{author}{\bibinfo{person}{Vladimir Kotlyar}, \bibinfo{person}{Keshav
  Pingali}, {and} \bibinfo{person}{Paul Stodghill}.}
  \bibinfo{year}{1997}\natexlab{b}.
\newblock \showarticletitle{A relational approach to the compilation of sparse
  matrix programs}. In \bibinfo{booktitle}{\emph{European Conference on
  Parallel Processing}}. Springer, \bibinfo{pages}{318--327}.
\newblock


\bibitem[Langr and Tvrdik(2016)]%
        {langr2016evaluation}
\bibfield{author}{\bibinfo{person}{Daniel Langr} {and} \bibinfo{person}{Pavel
  Tvrdik}.} \bibinfo{year}{2016}\natexlab{}.
\newblock \showarticletitle{Evaluation criteria for sparse matrix storage
  formats}.
\newblock \bibinfo{journal}{\emph{IEEE Transactions on parallel and distributed
  systems}} \bibinfo{volume}{27}, \bibinfo{number}{2} (\bibinfo{year}{2016}),
  \bibinfo{pages}{428--440}.
\newblock


\bibitem[Lattner et~al\mbox{.}(2020)]%
        {lattner2020mlir}
\bibfield{author}{\bibinfo{person}{Chris Lattner}, \bibinfo{person}{Mehdi
  Amini}, \bibinfo{person}{Uday Bondhugula}, \bibinfo{person}{Albert Cohen},
  \bibinfo{person}{Andy Davis}, \bibinfo{person}{Jacques Pienaar},
  \bibinfo{person}{River Riddle}, \bibinfo{person}{Tatiana Shpeisman},
  \bibinfo{person}{Nicolas Vasilache}, {and} \bibinfo{person}{Oleksandr
  Zinenko}.} \bibinfo{year}{2020}\natexlab{}.
\newblock \bibinfo{title}{MLIR: A Compiler Infrastructure for the End of
  Moore's Law}.
\newblock
\newblock
\showeprint[arxiv]{2002.11054}~[cs.PL]


\bibitem[Lefebvre and Feautrier(1998)]%
        {Lefebvre:1998:ASM}
\bibfield{author}{\bibinfo{person}{Vincent Lefebvre} {and}
  \bibinfo{person}{Paul Feautrier}.} \bibinfo{year}{1998}\natexlab{}.
\newblock \showarticletitle{Automatic storage management for parallel
  programs}.
\newblock \bibinfo{journal}{\emph{Parallel Comput.}} \bibinfo{volume}{24},
  \bibinfo{number}{3-4} (\bibinfo{date}{May} \bibinfo{year}{1998}),
  \bibinfo{pages}{649--671}.
\newblock


\bibitem[Li et~al\mbox{.}(2019)]%
        {reorder}
\bibfield{author}{\bibinfo{person}{Jiajia Li}, \bibinfo{person}{Bora U\c{c}ar},
  \bibinfo{person}{\"{U}mit~V. \c{C}ataly\"{u}rek}, \bibinfo{person}{Jimeng
  Sun}, \bibinfo{person}{Kevin Barker}, {and} \bibinfo{person}{Richard Vuduc}.}
  \bibinfo{year}{2019}\natexlab{}.
\newblock \showarticletitle{Efficient and Effective Sparse Tensor Reordering}.
  In \bibinfo{booktitle}{\emph{Proceedings of the ACM International Conference
  on Supercomputing}} (Phoenix, Arizona) \emph{(\bibinfo{series}{ICS '19})}.
  \bibinfo{publisher}{Association for Computing Machinery},
  \bibinfo{address}{New York, NY, USA}, \bibinfo{pages}{227–237}.
\newblock
\showISBNx{9781450360791}
\urldef\tempurl%
\url{https://doi.org/10.1145/3330345.3330366}
\showDOI{\tempurl}


\bibitem[Mateev et~al\mbox{.}(2000)]%
        {mateev2000}
\bibfield{author}{\bibinfo{person}{Nikolay Mateev}, \bibinfo{person}{Keshav
  Pingali}, \bibinfo{person}{Paul Stodghill}, {and} \bibinfo{person}{Vladimir
  Kotlyar}.} \bibinfo{year}{2000}\natexlab{}.
\newblock \showarticletitle{Next-Generation Generic Programming and Its
  Application to Sparse Matrix Computations}. In
  \bibinfo{booktitle}{\emph{Proceedings of the 14th International Conference on
  Supercomputing}} (Santa Fe, New Mexico, USA) \emph{(\bibinfo{series}{ICS
  '00})}. \bibinfo{publisher}{Association for Computing Machinery},
  \bibinfo{address}{New York, NY, USA}, \bibinfo{pages}{88–99}.
\newblock
\showISBNx{1581132700}
\urldef\tempurl%
\url{https://doi.org/10.1145/335231.335240}
\showDOI{\tempurl}


\bibitem[Mishra et~al\mbox{.}(2021)]%
        {mishra2021accelerating}
\bibfield{author}{\bibinfo{person}{Asit Mishra},
  \bibinfo{person}{Jorge~Albericio Latorre}, \bibinfo{person}{Jeff Pool},
  \bibinfo{person}{Darko Stosic}, \bibinfo{person}{Dusan Stosic},
  \bibinfo{person}{Ganesh Venkatesh}, \bibinfo{person}{Chong Yu}, {and}
  \bibinfo{person}{Paulius Micikevicius}.} \bibinfo{year}{2021}\natexlab{}.
\newblock \bibinfo{title}{Accelerating Sparse Deep Neural Networks}.
\newblock
\newblock
\showeprint[arxiv]{2104.08378}~[cs.LG]


\bibitem[Mohammadi et~al\mbox{.}(2019a)]%
        {mahdiarrayprop}
\bibfield{author}{\bibinfo{person}{Mahdi~Soltan Mohammadi},
  \bibinfo{person}{Kazem Cheshmi}, \bibinfo{person}{Maryam~Mehri Dehnavi},
  \bibinfo{person}{Anand Venkat}, \bibinfo{person}{Tomofumi Yuki}, {and}
  \bibinfo{person}{Michelle~Mills Strout}.} \bibinfo{year}{2019}\natexlab{a}.
\newblock \showarticletitle{Extending Index-Array Properties for Data
  Dependence Analysis}. In \bibinfo{booktitle}{\emph{Languages and Compilers
  for Parallel Computing}}, \bibfield{editor}{\bibinfo{person}{Mary Hall} {and}
  \bibinfo{person}{Hari Sundar}} (Eds.). \bibinfo{publisher}{Springer
  International Publishing}, \bibinfo{address}{Cham}, \bibinfo{pages}{78--93}.
\newblock
\showISBNx{978-3-030-34627-0}


\bibitem[Mohammadi et~al\mbox{.}(2019b)]%
        {mahdiinspector}
\bibfield{author}{\bibinfo{person}{Mahdi~Soltan Mohammadi},
  \bibinfo{person}{Tomofumi Yuki}, \bibinfo{person}{Kazem Cheshmi},
  \bibinfo{person}{Eddie~C. Davis}, \bibinfo{person}{Mary Hall},
  \bibinfo{person}{Maryam~Mehri Dehnavi}, \bibinfo{person}{Payal Nandy},
  \bibinfo{person}{Catherine Olschanowsky}, \bibinfo{person}{Anand Venkat},
  {and} \bibinfo{person}{Michelle~Mills Strout}.}
  \bibinfo{year}{2019}\natexlab{b}.
\newblock \showarticletitle{Sparse Computation Data Dependence Simplification
  for Efficient Compiler-Generated Inspectors}. In
  \bibinfo{booktitle}{\emph{Proceedings of the 40th ACM SIGPLAN Conference on
  Programming Language Design and Implementation}} (Phoenix, AZ, USA)
  \emph{(\bibinfo{series}{PLDI 2019})}. \bibinfo{publisher}{Association for
  Computing Machinery}, \bibinfo{address}{New York, NY, USA},
  \bibinfo{pages}{594–609}.
\newblock
\showISBNx{9781450367127}
\urldef\tempurl%
\url{https://doi.org/10.1145/3314221.3314646}
\showDOI{\tempurl}


\bibitem[Pugh and Shpeisman(1999)]%
        {sipr}
\bibfield{author}{\bibinfo{person}{William Pugh} {and} \bibinfo{person}{Tatiana
  Shpeisman}.} \bibinfo{year}{1999}\natexlab{}.
\newblock \showarticletitle{SIPR: A New Framework for Generating Efficient Code
  for Sparse Matrix Computations}. In \bibinfo{booktitle}{\emph{Languages and
  Compilers for Parallel Computing}},
  \bibfield{editor}{\bibinfo{person}{Siddhartha Chatterjee},
  \bibinfo{person}{Jan~F. Prins}, \bibinfo{person}{Larry Carter},
  \bibinfo{person}{Jeanne Ferrante}, \bibinfo{person}{Zhiyuan Li},
  \bibinfo{person}{David Sehr}, {and} \bibinfo{person}{Pen-Chung Yew}} (Eds.).
  \bibinfo{publisher}{Springer Berlin Heidelberg}, \bibinfo{address}{Berlin,
  Heidelberg}, \bibinfo{pages}{213--229}.
\newblock
\showISBNx{978-3-540-48319-9}


\bibitem[Quiller\'{e} and Rajopadhye(2000)]%
        {Polyhedral2000}
\bibfield{author}{\bibinfo{person}{Fabien Quiller\'{e}} {and}
  \bibinfo{person}{Sanjay Rajopadhye}.} \bibinfo{year}{2000}\natexlab{}.
\newblock \showarticletitle{Optimizing memory usage in the polyhedral model}.
\newblock \bibinfo{journal}{\emph{ACM Transactions on Programming Languages and
  Systems}} \bibinfo{volume}{22}, \bibinfo{number}{5} (\bibinfo{year}{2000}),
  \bibinfo{pages}{773--815}.
\newblock


\bibitem[Quilleré et~al\mbox{.}(2000)]%
        {quillere}
\bibfield{author}{\bibinfo{person}{Fabien Quilleré}, \bibinfo{person}{Sanjay
  Rajopadhye}, {and} \bibinfo{person}{Doran Wilde}.}
  \bibinfo{year}{2000}\natexlab{}.
\newblock \showarticletitle{Generation of Efficient Nested Loops from
  Polyhedra}.
\newblock \bibinfo{journal}{\emph{International journal of parallel
  programming}} \bibinfo{volume}{28}, \bibinfo{number}{5} (\bibinfo{date}{Oct}
  \bibinfo{year}{2000}), \bibinfo{pages}{469--498}.
\newblock
\showISSN{1573-7640}
\urldef\tempurl%
\url{https://doi.org/10.1023/A:1007554627716}
\showDOI{\tempurl}


\bibitem[Ramanujam and Sadayappan(1992)]%
        {ramanujam92}
\bibfield{author}{\bibinfo{person}{J. Ramanujam} {and} \bibinfo{person}{P.
  Sadayappan}.} \bibinfo{year}{1992}\natexlab{}.
\newblock \showarticletitle{Tiling of Iteration Spaces for Multicomputers}. In
  \bibinfo{booktitle}{\emph{Proceedings of the 1990 International Conference on
  Parallel Processing}}. \bibinfo{pages}{179--186}.
\newblock


\bibitem[Ricci and Levi-Civita(1900)]%
        {tensorindex}
\bibfield{author}{\bibinfo{person}{M.~M.~G. Ricci} {and} \bibinfo{person}{T.
  Levi-Civita}.} \bibinfo{year}{1900}\natexlab{}.
\newblock \showarticletitle{Méthodes de calcul différentiel absolu et leurs
  applications}.
\newblock \bibinfo{journal}{\emph{Math. Ann.}} \bibinfo{volume}{54},
  \bibinfo{number}{1} (\bibinfo{date}{Mar} \bibinfo{year}{1900}),
  \bibinfo{pages}{125--201}.
\newblock
\showISSN{1432-1807}
\urldef\tempurl%
\url{https://doi.org/10.1007/BF01454201}
\showDOI{\tempurl}


\bibitem[{Rong} et~al\mbox{.}(2016)]%
        {sparso}
\bibfield{author}{\bibinfo{person}{H. {Rong}}, \bibinfo{person}{J. {Park}},
  \bibinfo{person}{L. {Xiang}}, \bibinfo{person}{T.~A. {Anderson}}, {and}
  \bibinfo{person}{M. {Smelyanskiy}}.} \bibinfo{year}{2016}\natexlab{}.
\newblock \showarticletitle{Sparso: Context-driven optimizations of sparse
  linear algebra}. In \bibinfo{booktitle}{\emph{2016 International Conference
  on Parallel Architecture and Compilation Techniques (PACT)}}.
  \bibinfo{pages}{247--259}.
\newblock
\urldef\tempurl%
\url{https://doi.org/10.1145/2967938.2967943}
\showDOI{\tempurl}


\bibitem[Saad(2003)]%
        {saad2003}
\bibfield{author}{\bibinfo{person}{Yousef Saad}.}
  \bibinfo{year}{2003}\natexlab{}.
\newblock \bibinfo{booktitle}{\emph{Iterative Methods for Sparse Linear
  Systems} (\bibinfo{edition}{second} ed.)}.
\newblock \bibinfo{publisher}{Society for Industrial and Applied Mathematics}.
\newblock
\urldef\tempurl%
\url{https://doi.org/10.1137/1.9780898718003}
\showDOI{\tempurl}
\showeprint{https://epubs.siam.org/doi/pdf/10.1137/1.9780898718003}


\bibitem[Senanayake et~al\mbox{.}(2020)]%
        {Senanayake2020}
\bibfield{author}{\bibinfo{person}{Ryan Senanayake}, \bibinfo{person}{Changwan
  Hong}, \bibinfo{person}{Ziheng Wang}, \bibinfo{person}{Amalee Wilson},
  \bibinfo{person}{Stephen Chou}, \bibinfo{person}{Shoaib Kamil},
  \bibinfo{person}{Saman Amarasinghe}, {and} \bibinfo{person}{Fredrik
  Kjolstad}.} \bibinfo{year}{2020}\natexlab{}.
\newblock \showarticletitle{A Sparse Iteration Space Transformation Framework
  for Sparse Tensor Algebra}. In \bibinfo{booktitle}{\emph{OOPSLA}}.
\newblock


\bibitem[Shen and Wonnacott(1998)]%
        {Shen98}
\bibfield{author}{\bibinfo{person}{Tina Shen} {and} \bibinfo{person}{David
  Wonnacott}.} \bibinfo{year}{1998}\natexlab{}.
\newblock \showarticletitle{Automatic Memory Remapping for Time Skewing}. In
  \bibinfo{booktitle}{\emph{Mid-Atlantic Student Workshop on Programming
  Languages and Systems (MASPLAS)}}.
\newblock


\bibitem[Sidiropoulos et~al\mbox{.}(2017)]%
        {Sidiropoulos:2017:survey}
\bibfield{author}{\bibinfo{person}{N.~D. Sidiropoulos}, \bibinfo{person}{L. {De
  Lathauwer}}, \bibinfo{person}{X. Fu}, \bibinfo{person}{K. Huang},
  \bibinfo{person}{E.~E. Papalexakis}, {and} \bibinfo{person}{C. Faloutsos}.}
  \bibinfo{year}{2017}\natexlab{}.
\newblock \showarticletitle{Tensor Decomposition for Signal Processing and
  Machine Learning}.
\newblock \bibinfo{journal}{\emph{IEEE Transactions on Signal Processing}}
  \bibinfo{volume}{65}, \bibinfo{number}{13} (\bibinfo{date}{July}
  \bibinfo{year}{2017}), \bibinfo{pages}{3551--3582}.
\newblock
\showISSN{1053-587X}
\urldef\tempurl%
\url{https://doi.org/10.1109/TSP.2017.2690524}
\showDOI{\tempurl}


\bibitem[Smith et~al\mbox{.}(2017)]%
        {frosttdataset}
\bibfield{author}{\bibinfo{person}{Shaden Smith}, \bibinfo{person}{Jee~W.
  Choi}, \bibinfo{person}{Jiajia Li}, \bibinfo{person}{Richard Vuduc},
  \bibinfo{person}{Jongsoo Park}, \bibinfo{person}{Xing Liu}, {and}
  \bibinfo{person}{George Karypis}.} \bibinfo{year}{2017}\natexlab{}.
\newblock \bibinfo{booktitle}{\emph{{FROSTT}: The Formidable Repository of Open
  Sparse Tensors and Tools}}.
\newblock
\urldef\tempurl%
\url{http://frostt.io/}
\showURL{%
\tempurl}


\bibitem[Smith et~al\mbox{.}(2015)]%
        {csf}
\bibfield{author}{\bibinfo{person}{Shaden Smith}, \bibinfo{person}{Niranjay
  Ravindran}, \bibinfo{person}{Nicholas~D. Sidiropoulos}, {and}
  \bibinfo{person}{George Karypis}.} \bibinfo{year}{2015}\natexlab{}.
\newblock \showarticletitle{SPLATT: Efficient and Parallel Sparse Tensor-Matrix
  Multiplication}. In \bibinfo{booktitle}{\emph{2015 IEEE International
  Parallel and Distributed Processing Symposium}}. \bibinfo{pages}{61--70}.
\newblock
\urldef\tempurl%
\url{https://doi.org/10.1109/IPDPS.2015.27}
\showDOI{\tempurl}


\bibitem[Solar-Lezama et~al\mbox{.}(2006)]%
        {lezama2006sketching}
\bibfield{author}{\bibinfo{person}{Armando Solar-Lezama},
  \bibinfo{person}{Liviu Tancau}, \bibinfo{person}{Rastislav Bodik},
  \bibinfo{person}{Sanjit Seshia}, {and} \bibinfo{person}{Vijay Saraswat}.}
  \bibinfo{year}{2006}\natexlab{}.
\newblock \showarticletitle{Combinatorial Sketching for Finite Programs}. In
  \bibinfo{booktitle}{\emph{Proceedings of the 12th International Conference on
  Architectural Support for Programming Languages and Operating Systems}} (San
  Jose, California, USA) \emph{(\bibinfo{series}{ASPLOS XII})}.
  \bibinfo{publisher}{Association for Computing Machinery},
  \bibinfo{address}{New York, NY, USA}, \bibinfo{pages}{404–415}.
\newblock
\showISBNx{1595934510}
\urldef\tempurl%
\url{https://doi.org/10.1145/1168857.1168907}
\showDOI{\tempurl}


\bibitem[Stodghill(1997)]%
        {stodghill1997thesis}
\bibfield{author}{\bibinfo{person}{Paul~Vinson Stodghill}.}
  \bibinfo{year}{1997}\natexlab{}.
\newblock \emph{\bibinfo{title}{A Relational Approach to the Automatic
  Generation of Sequential Sparse Matrix Codes}}.
\newblock \bibinfo{thesistype}{Ph.\,D. Dissertation}. \bibinfo{school}{Cornell
  University}, \bibinfo{address}{USA}.
\newblock
\newblock
\shownote{UMI Order No. GAX97-33644}.


\bibitem[Strout et~al\mbox{.}(2003)]%
        {StroutPLDI03}
\bibfield{author}{\bibinfo{person}{Michelle~Mills Strout},
  \bibinfo{person}{Larry Carter}, {and} \bibinfo{person}{Jeanne Ferrante}.}
  \bibinfo{year}{2003}\natexlab{}.
\newblock \showarticletitle{Compile-time Composition of Run-time Data and
  Iteration Reorderings}. In \bibinfo{booktitle}{\emph{Proceedings of the {ACM}
  {SIGPLAN} Conference on Programming Language Design and Implementation
  (PLDI)}}. \bibinfo{publisher}{ACM}, \bibinfo{address}{New York, NY, USA}.
\newblock


\bibitem[Strout et~al\mbox{.}(1998)]%
        {StroutEtAl98}
\bibfield{author}{\bibinfo{person}{Michelle~Mills Strout},
  \bibinfo{person}{Larry Carter}, \bibinfo{person}{Jeanne Ferrante}, {and}
  \bibinfo{person}{Beth Simon}.} \bibinfo{year}{1998}\natexlab{}.
\newblock \showarticletitle{Schedule-Independent Storage Mapping for Loops}. In
  \bibinfo{booktitle}{\emph{Proceedings of the Eighth International Conference
  on Architectural Support for Programming Languages and Operating Systems}}.
  \bibinfo{address}{San Jose, California}, \bibinfo{pages}{24--33}.
\newblock


\bibitem[Strout et~al\mbox{.}(2018)]%
        {Strout18}
\bibfield{author}{\bibinfo{person}{M.~M. Strout}, \bibinfo{person}{M. Hall},
  {and} \bibinfo{person}{C. Olschanowsky}.} \bibinfo{year}{2018}\natexlab{}.
\newblock \showarticletitle{The Sparse Polyhedral Framework: Composing
  Compiler-Generated Inspector-Executor Code}.
\newblock \bibinfo{journal}{\emph{Proc. IEEE}} \bibinfo{volume}{106},
  \bibinfo{number}{11} (\bibinfo{date}{Nov} \bibinfo{year}{2018}),
  \bibinfo{pages}{1921--1934}.
\newblock


\bibitem[Strout et~al\mbox{.}(2016)]%
        {Strout16}
\bibfield{author}{\bibinfo{person}{Michelle~Mills Strout},
  \bibinfo{person}{Alan LaMielle}, \bibinfo{person}{Larry Carter},
  \bibinfo{person}{Jeanne Ferrante}, \bibinfo{person}{Barbara Kreaseck}, {and}
  \bibinfo{person}{Catherine Olschanowsky}.} \bibinfo{year}{2016}\natexlab{}.
\newblock \showarticletitle{An Approach for Code Generation in the Sparse
  Polyhedral Framework}.
\newblock \bibinfo{journal}{\emph{Parallel Comput.}} \bibinfo{volume}{53},
  \bibinfo{number}{C} (\bibinfo{date}{April} \bibinfo{year}{2016}),
  \bibinfo{pages}{32--57}.
\newblock


\bibitem[Thies et~al\mbox{.}(2007)]%
        {Thies2007}
\bibfield{author}{\bibinfo{person}{William Thies},
  \bibinfo{person}{Fr\'{e}d\'{e}ric Vivien}, {and} \bibinfo{person}{Saman
  Amarasinghe}.} \bibinfo{year}{2007}\natexlab{}.
\newblock \showarticletitle{A step towards unifying schedule and storage
  optimization}.
\newblock \bibinfo{journal}{\emph{ACM Transactions on Programming Languages and
  Systems}} \bibinfo{volume}{29}, \bibinfo{number}{6} (\bibinfo{year}{2007}),
  \bibinfo{pages}{34}.
\newblock


\bibitem[Tian et~al\mbox{.}(2021)]%
        {comet}
\bibfield{author}{\bibinfo{person}{Ruiqin Tian}, \bibinfo{person}{Luanzheng
  Guo}, \bibinfo{person}{Jiajia Li}, \bibinfo{person}{Bin Ren}, {and}
  \bibinfo{person}{Gokcen Kestor}.} \bibinfo{year}{2021}\natexlab{}.
\newblock \showarticletitle{A High Performance Sparse Tensor Algebra Compiler
  in MLIR}. In \bibinfo{booktitle}{\emph{The Seventh Annual Workshop on the
  LLVM Compiler Infrastructure in HPC}}.
\newblock


\bibitem[Valiev et~al\mbox{.}(2010)]%
        {NWChem}
\bibfield{author}{\bibinfo{person}{M. Valiev}, \bibinfo{person}{E.J. Bylaska},
  \bibinfo{person}{N. Govind}, \bibinfo{person}{K. Kowalski},
  \bibinfo{person}{T.P. Straatsma}, \bibinfo{person}{H.J.J. {Van Dam}},
  \bibinfo{person}{D. Wang}, \bibinfo{person}{J. Nieplocha},
  \bibinfo{person}{E. Apra}, \bibinfo{person}{T.L. Windus}, {and}
  \bibinfo{person}{W.A. {de Jong}}.} \bibinfo{year}{2010}\natexlab{}.
\newblock \showarticletitle{NWChem: A comprehensive and scalable open-source
  solution for large scale molecular simulations}.
\newblock \bibinfo{journal}{\emph{Computer Physics Communications}}
  \bibinfo{volume}{181}, \bibinfo{number}{9} (\bibinfo{year}{2010}),
  \bibinfo{pages}{1477--1489}.
\newblock
\showISSN{0010-4655}
\urldef\tempurl%
\url{https://doi.org/10.1016/j.cpc.2010.04.018}
\showDOI{\tempurl}


\bibitem[van~der Spek and Wijshoff(2011)]%
        {sublimation}
\bibfield{author}{\bibinfo{person}{Harmen L.~A. van~der Spek} {and}
  \bibinfo{person}{Harry A.~G. Wijshoff}.} \bibinfo{year}{2011}\natexlab{}.
\newblock \showarticletitle{Sublimation: Expanding Data Structures to Enable
  Data Instance Specific Optimizations}. In \bibinfo{booktitle}{\emph{Languages
  and Compilers for Parallel Computing}},
  \bibfield{editor}{\bibinfo{person}{Keith Cooper}, \bibinfo{person}{John
  Mellor-Crummey}, {and} \bibinfo{person}{Vivek Sarkar}} (Eds.).
  \bibinfo{publisher}{Springer Berlin Heidelberg}, \bibinfo{address}{Berlin,
  Heidelberg}, \bibinfo{pages}{106--120}.
\newblock
\showISBNx{978-3-642-19595-2}


\bibitem[Vasilache et~al\mbox{.}(2006)]%
        {violateddependence}
\bibfield{author}{\bibinfo{person}{Nicolas Vasilache}, \bibinfo{person}{Cedric
  Bastoul}, \bibinfo{person}{Albert Cohen}, {and} \bibinfo{person}{Sylvain
  Girbal}.} \bibinfo{year}{2006}\natexlab{}.
\newblock \showarticletitle{Violated Dependence Analysis}. In
  \bibinfo{booktitle}{\emph{Proceedings of the 20th Annual International
  Conference on Supercomputing}} (Cairns, Queensland, Australia)
  \emph{(\bibinfo{series}{ICS '06})}. \bibinfo{publisher}{Association for
  Computing Machinery}, \bibinfo{address}{New York, NY, USA},
  \bibinfo{pages}{335–344}.
\newblock
\showISBNx{1595932828}
\urldef\tempurl%
\url{https://doi.org/10.1145/1183401.1183448}
\showDOI{\tempurl}


\bibitem[Venkat et~al\mbox{.}(2015)]%
        {anandpldi}
\bibfield{author}{\bibinfo{person}{Anand Venkat}, \bibinfo{person}{Mary Hall},
  {and} \bibinfo{person}{Michelle Strout}.} \bibinfo{year}{2015}\natexlab{}.
\newblock \showarticletitle{Loop and Data Transformations for Sparse Matrix
  Code}. In \bibinfo{booktitle}{\emph{Proceedings of the 36th ACM SIGPLAN
  Conference on Programming Language Design and Implementation}} (Portland, OR,
  USA) \emph{(\bibinfo{series}{PLDI '15})}. \bibinfo{publisher}{Association for
  Computing Machinery}, \bibinfo{address}{New York, NY, USA},
  \bibinfo{pages}{521–532}.
\newblock
\showISBNx{9781450334686}
\urldef\tempurl%
\url{https://doi.org/10.1145/2737924.2738003}
\showDOI{\tempurl}


\bibitem[Venkat et~al\mbox{.}(2016)]%
        {wavefront}
\bibfield{author}{\bibinfo{person}{Anand Venkat}, \bibinfo{person}{Mahdi~Soltan
  Mohammadi}, \bibinfo{person}{Jongsoo Park}, \bibinfo{person}{Hongbo Rong},
  \bibinfo{person}{Rajkishore Barik}, \bibinfo{person}{Michelle~Mills Strout},
  {and} \bibinfo{person}{Mary Hall}.} \bibinfo{year}{2016}\natexlab{}.
\newblock \showarticletitle{Automating Wavefront Parallelization for Sparse
  Matrix Computations}. In \bibinfo{booktitle}{\emph{Proceedings of the
  International Conference for High Performance Computing, Networking, Storage
  and Analysis}} (Salt Lake City, Utah) \emph{(\bibinfo{series}{SC '16})}.
  \bibinfo{publisher}{IEEE Press}, Article \bibinfo{articleno}{41},
  \bibinfo{numpages}{12}~pages.
\newblock
\showISBNx{9781467388153}


\bibitem[Virtanen et~al\mbox{.}(2020)]%
        {2020SciPy-NMeth}
\bibfield{author}{\bibinfo{person}{Pauli Virtanen}, \bibinfo{person}{Ralf
  Gommers}, \bibinfo{person}{Travis~E. Oliphant}, \bibinfo{person}{Matt
  Haberland}, \bibinfo{person}{Tyler Reddy}, \bibinfo{person}{David
  Cournapeau}, \bibinfo{person}{Evgeni Burovski}, \bibinfo{person}{Pearu
  Peterson}, \bibinfo{person}{Warren Weckesser}, \bibinfo{person}{Jonathan
  Bright}, \bibinfo{person}{St{\'e}fan~J. {van der Walt}},
  \bibinfo{person}{Matthew Brett}, \bibinfo{person}{Joshua Wilson},
  \bibinfo{person}{K.~Jarrod Millman}, \bibinfo{person}{Nikolay Mayorov},
  \bibinfo{person}{Andrew R.~J. Nelson}, \bibinfo{person}{Eric Jones},
  \bibinfo{person}{Robert Kern}, \bibinfo{person}{Eric Larson},
  \bibinfo{person}{C~J Carey}, \bibinfo{person}{{\.I}lhan Polat},
  \bibinfo{person}{Yu Feng}, \bibinfo{person}{Eric~W. Moore},
  \bibinfo{person}{Jake {VanderPlas}}, \bibinfo{person}{Denis Laxalde},
  \bibinfo{person}{Josef Perktold}, \bibinfo{person}{Robert Cimrman},
  \bibinfo{person}{Ian Henriksen}, \bibinfo{person}{E.~A. Quintero},
  \bibinfo{person}{Charles~R. Harris}, \bibinfo{person}{Anne~M. Archibald},
  \bibinfo{person}{Ant{\^o}nio~H. Ribeiro}, \bibinfo{person}{Fabian Pedregosa},
  \bibinfo{person}{Paul {van Mulbregt}}, {and} \bibinfo{person}{{SciPy 1.0
  Contributors}}.} \bibinfo{year}{2020}\natexlab{}.
\newblock \showarticletitle{{{SciPy} 1.0: Fundamental Algorithms for Scientific
  Computing in Python}}.
\newblock \bibinfo{journal}{\emph{Nature Methods}}  \bibinfo{volume}{17}
  (\bibinfo{year}{2020}), \bibinfo{pages}{261--272}.
\newblock
\urldef\tempurl%
\url{https://doi.org/10.1038/s41592-019-0686-2}
\showDOI{\tempurl}


\bibitem[Wang et~al\mbox{.}(2019)]%
        {DBLP:journals/corr/abs-1909-01315}
\bibfield{author}{\bibinfo{person}{Minjie Wang}, \bibinfo{person}{Lingfan Yu},
  \bibinfo{person}{Da Zheng}, \bibinfo{person}{Quan Gan}, \bibinfo{person}{Yu
  Gai}, \bibinfo{person}{Zihao Ye}, \bibinfo{person}{Mufei Li},
  \bibinfo{person}{Jinjing Zhou}, \bibinfo{person}{Qi Huang},
  \bibinfo{person}{Chao Ma}, \bibinfo{person}{Ziyue Huang},
  \bibinfo{person}{Qipeng Guo}, \bibinfo{person}{Hao Zhang},
  \bibinfo{person}{Haibin Lin}, \bibinfo{person}{Junbo Zhao},
  \bibinfo{person}{Jinyang Li}, \bibinfo{person}{Alexander~J. Smola}, {and}
  \bibinfo{person}{Zheng Zhang}.} \bibinfo{year}{2019}\natexlab{}.
\newblock \showarticletitle{Deep Graph Library: Towards Efficient and Scalable
  Deep Learning on Graphs}.
\newblock \bibinfo{journal}{\emph{CoRR}}  \bibinfo{volume}{abs/1909.01315}
  (\bibinfo{year}{2019}).
\newblock
\showeprint[arXiv]{1909.01315}
\urldef\tempurl%
\url{http://arxiv.org/abs/1909.01315}
\showURL{%
\tempurl}


\bibitem[Wolf and Lam(1991)]%
        {wolflam91}
\bibfield{author}{\bibinfo{person}{Michael~E. Wolf} {and}
  \bibinfo{person}{Monica~S. Lam}.} \bibinfo{year}{1991}\natexlab{}.
\newblock \showarticletitle{A Data Locality Optimizing Algorithm}. In
  \bibinfo{booktitle}{\emph{Proceedings of the ACM SIGPLAN 1991 Conference on
  Programming Language Design and Implementation}} (Toronto, Ontario, Canada)
  \emph{(\bibinfo{series}{PLDI '91})}. \bibinfo{publisher}{Association for
  Computing Machinery}, \bibinfo{address}{New York, NY, USA},
  \bibinfo{pages}{30–44}.
\newblock
\showISBNx{0897914287}
\urldef\tempurl%
\url{https://doi.org/10.1145/113445.113449}
\showDOI{\tempurl}


\bibitem[Wolfe(1989)]%
        {wolfe89}
\bibfield{author}{\bibinfo{person}{M. Wolfe}.} \bibinfo{year}{1989}\natexlab{}.
\newblock \showarticletitle{More iteration space tiling}. In
  \bibinfo{booktitle}{\emph{Supercomputing '89:Proceedings of the 1989 ACM/IEEE
  Conference on Supercomputing}}. \bibinfo{pages}{655--664}.
\newblock
\urldef\tempurl%
\url{https://doi.org/10.1145/76263.76337}
\showDOI{\tempurl}


\bibitem[Zhao et~al\mbox{.}(2018)]%
        {zhaocc18}
\bibfield{author}{\bibinfo{person}{Jie Zhao}, \bibinfo{person}{Michael Kruse},
  {and} \bibinfo{person}{Albert Cohen}.} \bibinfo{year}{2018}\natexlab{}.
\newblock \showarticletitle{A Polyhedral Compilation Framework for Loops with
  Dynamic Data-Dependent Bounds}. In \bibinfo{booktitle}{\emph{Proceedings of
  the 27th International Conference on Compiler Construction}} (Vienna,
  Austria) \emph{(\bibinfo{series}{CC 2018})}. \bibinfo{publisher}{Association
  for Computing Machinery}, \bibinfo{address}{New York, NY, USA},
  \bibinfo{pages}{14–24}.
\newblock
\showISBNx{9781450356442}
\urldef\tempurl%
\url{https://doi.org/10.1145/3178372.3179509}
\showDOI{\tempurl}


\end{thebibliography}

\appendix

\pagebreak

\section{Synthesizing find algorithms}
\label{app:synth}

This section describes the details of the synthesis process used in augmented polyhedral scanning shown on Line~\ref{line:check}-\ref{line:generate} in Fig.~\ref{alg:codegen}. The goal of the synthesis is to consume the bounds and conditions that arise in the program and index array properties specified to decide if implementing the loop as a find algorithm is possible and determine the proper template arguments to generate.

We will demonstrate this process of code synthesis on the running example of the sparse vector dot product shown in Fig.~\ref{fig:sparse-dot-all}. Polyhedral scanning in Algorithm~\ref{alg:codegen} will produce the loop bounds and conditions as shown in Table~\ref{tab:scan}. The synthesis will work entirely with this set of bounds and conditions without further interactions with the polyhedral framework.

\begin{table}[t]
    \centering
        \caption{Polyhedral scanning result for sparse dot product, Fig.~\ref{fig:sparse-is}.}
    \begin{tabular}{c|c|c|c} \hline
        $i$ & $L_i$ & $\mathit{UF}$ & $\mathit{C_{i}}$ \\ \hline \hline
        $pA$ & $0 \leq pA < A.len$ & $A.idx$ & \\ \hline
        $i$ & $i = A.idx(pA)$ & & \\ \hline
        $pB$ & $0 \leq pB < B.len$ & $B.idx$ & $ B.idx(pB) = i$ \\ \hline
    \end{tabular}

    \label{tab:scan}
\end{table}

\subsection{Theory components}

SMT solvers take a formula in first-order logic, a \emph{theory}, and determine if the system is \emph{satisfiable}, that is if there exist assignments to the variables so that the formula evaluates to true. It is a generalization to Boolean SAT that incorporates higher-level abstractions such as integer arithmetic and arrays. Note also that in some general cases not encountered in this work, Presburger formula with uninterpreted functions are undecidable~\cite{whats}, regardless of whether SMT solvers or polyhedral frameworks are used.

The theory to prove consists of 3 components: 1) the bounds and conditions generated through polyhedral scanning, 2) adapted sets of index array properties. 3) assumptions of the algorithms, including the generated template arguments.

\textbf{Bounds and conditions}
\begin{itemize}
    \item \textbf{C1}: \emph{Lexicographical ordering of containing loop indices.} This is formed by using two sets of indices representing two iteration instances. For example, the following formula denotes the lexicographical ordering of loop $pA$.
    $$pA < pA'$$
    \item \textbf{C2}: \emph{Loop bounds.} Note that we use two versions representing the two loop instances. For the $pA$ loop,
    $$0 \leq pA < A.len \wedge 0 \leq pA' < A.len$$
    \item \textbf{C3}: \emph{Other equality/inequality conditions.} For the condition that $A.idx(pA)$ denotes the coordinate of nonzero at $pA$ in $A$, this is expressed as follows.
    $$i = A.idx(pA) \wedge i' = A.idx(pA')$$
    \item \textbf{C4}: \emph{Find condition.} This represent the condition used for the find expression.
    $$B.idx(pB) = i \wedge B.idx(pB') = i'$$
\end{itemize}

\textbf{Index array properties} contains two parts: the index array properties used by outer loops and the index array property that we are investigating. This separation is used to differentiate the treatments of these two components during the synthesis.

\begin{itemize}

\item \textbf{P5}: \emph{Applied} properties are generated by substituting the two set of loop indices into the property formula. 
$$pA < pA' \rightarrow A.idx(pA) < A.idx(pA') \wedge pA' < pA \rightarrow A.idx(pA') < A.idx(pA)$$

\item \textbf{P6}: \emph{Investigating} properties are generated similarly based on how the index array is used in the program.
$$pB < pB' \rightarrow B.idx(pB) < B.idx(pB') \wedge pB' < pB \rightarrow B.idx(pB') < A.idx(pB)$$

\end{itemize}

\textbf{Algorithm components} are intrinsic to the algorithms and the template arguments. Here we present a general type of algorithm assumption and three assumptions that are specific to the sequential iteration algorithm.

\begin{itemize}

\item \textbf{A7}: \emph{State indices.} (Section~\ref{sec:state}) This is a general condition on the saved state of the algorithm, applied whenever the algorithm uses persistence storage. For example, the sequential iteration algorithm saves the last iterated position in $pB$ variable initialized at the top of the loops in Fig.~\ref{fig:sparse-dot-opt}. This will have a state indices condition of $True$. If the state index is defined for each $pA$ loop then this condition will be $pA = pA'$.

\item \textbf{A8}: \texttt{Loop order (specific).} This is influenced by the template arguments defining in which direction (increasing/decreasing) $pB$ will move in the generated code. For increasing iteration:
$$pB < pB'$$

\item \textbf{A9}: \texttt{Stop condition (specific).} This represents the comparison expression \texttt{A0 > B.idx[pB]} in the code. This condition is also an argument to the template and is generated through construction.
$$pB < pB'$$

\item \textbf{A10}: \texttt{Exclusivity (specific).} This is added to force the find condition to be different and not overlapping. Combined with the loop order condition, this forces an order in which the find condition is allowed to change. This is especially important for blocked data structures: an outer loop may iterate over the blocks and an inner loop to find inside the block. When this is placed on the inner loop, then the outer loop can only iterate the blocks according to how the inner loop is iterating.
$$\neg(i = B.idx(pB')) \wedge \neg(i' = B.idx(pB))$$
\end{itemize}

Note that \textbf{C1-3} and \textbf{P5} indicates the context of the outer loops that has been generated. We can call this set of theory components the context, \textbf{C}.
$$C = C1 \wedge C2 \wedge C3 \wedge P5$$

\subsection{Deciding if Sequential Iteration can be Applied}

While we describe only the process for sequential iteration, the process is similar for both the HashMap algorithm and algorithms like Binary Search. Algorithm~\ref{alg:determine} shows the pseudo-code for this code generation process. The assumptions of the algorithms are designed to test if the algorithm produces correct results under the conditions from the program and the generated template arguments. For sequential iteration, we extracted two such assumptions from the algorithm.

The \textbf{first assumption} is to prove that, for all pairs of values to find during the program's execution, they follow the index array properties of the UF to be inverted
($P5$). Let $\mathbf{i}$ and $\mathbf{i'}$ represent the two integer set variables of the two iteration instances. This assumption can be encoded as Equation~\ref{eq:must-forward}. Note the left-hand side of the arrow represents the outer loops' context and conditions generated with template arguments.

\begin{equation}
    \forall \mathbf{i}, \mathbf{i'} (C \wedge C4 \wedge A7 \wedge A8 \wedge A10) \rightarrow \neg P6
    \label{eq:must-forward}
\end{equation}

Based on logical transformations to prove Equation~\ref{eq:must-forward} always holds for all indices values is to prove Equation~\ref{eq:must} \emph{unsatisfiable}.

\begin{equation}
    C \wedge C4 \wedge A7 \wedge A8 \wedge A10 \wedge \neg P6
    \label{eq:must}
\end{equation}

The \textbf{second assumption} uses the generated stop condition in the template arguments. This conditional expression is represented as $A9$. 
This assumption is satisfied when Equation~\ref{eq:imp} is proven \emph{unsatisfiable}. This condition states that it is impossible to find further solutions after the stop condition is triggered. $C4$ is not included here because it is irrelevant: The stop condition is designed to capture how values are changing based on the context and arguments; Whether the context and generated arguments can produce a match is captured by the first assumption.

\begin{equation}
    C \wedge P6 \wedge A7 \wedge A8 \wedge A9
    \label{eq:imp}
\end{equation}

\begin{algorithm}[t]
\SetAlgoLined
 \KwOut{$apply$: if this algorithm applies. $params$: parameters to the template.}
 $apply = false$; $curBestParams = \emptyset$\;
 \For {Enumerated template arguments: \emph{Loop order}}{
 \For {Constructed template arguments: \emph{State indices, stop condition}}{
    Refine construction\;
    \If{Assumptions failed to satisfy \label{line:test-assumptions}}{
    Roll back construction\;
    \textbf{Break}\;
    }
 }
 \If{Current set of parameters results in lower algorithmic complexity}{
    $curBestParams = Enumerated + Constructed$\;
}
 }
 \Return $apply$, $curBestParams$\;
 \caption{Checking if \emph{sequential iteration} applies.}
 \label{alg:determine}
\end{algorithm}

\subsection{State indices}
\label{sec:state}

State indices, $S$, is the central abstraction we introduced to encapsulate the persistent storage needed by a specific find algorithm. For an algorithm, the state indices $S$ is a subset of the containing loops' indices.
As in the example of sparse dot produce, the set of the state indices determines where $pB$ is initialized: outside or inside the $pA$ loop. We first produce the assumptions of state indices by considering the implementations shown in Fig.~\ref{fig:sparse-dot-state}.

\begin{figure*}[t]
\centering
\begin{subfigure}[t]{0.8\textwidth}
\centering
\begin{tabular}{c}
\begin{lstlisting}[language=C++,escapechar=|]
pB = 0;
for (int pA = 0; pA < A.len; ++pA) {
 i = A.idx[pA];
 while (pB < B.len && i > B.idx[pB]) ++pB;
 if (pB < B.len && i == B.idx[pB])
   { v += A.val[pA] * B.val[pB]; ++pB; }
}
\end{lstlisting}
\end{tabular}
\caption{Optimized code (reproduced Fig.~\ref{fig:sparse-dot-opt}). $S = \emptyset$.}
\label{fig:sparse-dot-opt-rep}
\end{subfigure}

\begin{subfigure}[t]{0.45\textwidth}
\centering
\begin{tabular}{c}
\begin{lstlisting}[language=C++,escapechar=|]


for (int pA=0; pA<A.len; ++pA) {
 pB = 0;
 i = A.idx[pA];
 while (pB<B.len && i>B.idx[pB]) 
  ++pB;
 if (pB<B.len && i==B.idx[pB]) {
 
  v+=A.val[pA]*B.val[pB]; 
  ++pB; }
}
\end{lstlisting}
\end{tabular}
\caption{Sub-optimal state initialization. $S = \{pA\}$.}
\label{fig:sparse-dot-dup}
\end{subfigure}
~
\begin{subfigure}[t]{0.45\textwidth}
\centering
\begin{tabular}{c}
\begin{lstlisting}[language=C++,escapechar=|]
for (int pA=0; pA<A.len; ++pA)
 pB[pA] = 0;
for (int pA=0; pA<A.len; ++pA) {

 i = A.idx[pA];
 while (pB[pA]<B.len && 
  i>B.idx[pB[pA]]) ++pB[pA];
 if (pB[pA]<B.len && 
     i==B.idx[pB[pA]]) {
  v+=A.val[pA]*B.val[pB[pA]]; 
  ++pB; }
}
\end{lstlisting}
\end{tabular}
\caption{Equivalent sub-optimal state initialization.}
\label{fig:sparse-dot-dup-equiv}
\end{subfigure}

\caption{Dot product between sparse vectors.}

\label{fig:sparse-dot-state}
\end{figure*}

From Fig.~\ref{fig:sparse-dot-state} we can deduce two assumptions of state indices. 1. If state indices $S^*$ produces correct code (Fig.~\ref{fig:sparse-dot-opt-rep}: $S^* = \emptyset$), adding any index to this set will also produce correct code (Fig.~\ref{fig:sparse-dot-dup}: $S = S^* \cup \{pA\}$). 2. When an index is part of the state indices a distinct state variable is attached to the loop variable's value: compare the logically equivalent implementations of Fig.~\ref{fig:sparse-dot-dup} and Fig.~\ref{fig:sparse-dot-dup-equiv}. 

The second assumption describes the state indices as a value-based system, directly contributing how $A7$ is constructed as an equality relation. Consider Fig.~\ref{fig:sparse-dot-dup-equiv}($S = \{pA\}$) and $pA = pA'$, the two loop instances will reference $pB[pA]$ and $pB[pA']$ which will point to the same state variable.

After introducing $A7$ into the logical formula, we can determine $S$ based a greedy algorithm produced by the first assumption: Starting from the full set of loop indices and we can eliminate indices one by one from the inner indices to outer indices until no indices can be eliminated without failing the assumptions of the algorithm. The solution is unique for most cases and will be reached by this algorithm. There are cases when the solution is not unique due to one loop indices capturing multiple aspects of the tensor and multiple loop indices capturing the same aspect of the tensor. For example, in the loop over nonzeros in COO, $nnz$, captures both the row and column indices of nonzeros; another loop may exist to match only with the row of COO,$row$. Thus if a find picks the row of nonzero as part of the state indices, $nnz$ and $r$ are both valid solutions. To solve this ambiguous situation, we further introduced \emph{reduced indices} as a possible choice for state indices that comes from generated non-loop set variables of the iteration space. This variable can capture partial meaning from iterators. Simply, with reduced indices, we are able to automatically deduce that $nnz = col\times row$, making $row$ a better pick over $nnz$ because it sheds unnecessary information.
Algorithm~\ref{alg:state} is the pseudo-code for constructing the set of state indices.

\begin{algorithm}[t]
\SetAlgoLined
\KwIn{$M$: The set of containing loop and reduced indices.}
\KwOut{The set of state indices, $S$}
 $S$ = $M$\;
 \For{Index $i \in M$ (innermost loop first, loop index before reduced indices)}{
   $S = S -\{i\}$\;
   $A7 = \bigwedge_{j \in S}(j = j')$\;
   \uIf{Failed the assumption test (Line~\ref{line:test-assumptions} of Algorithm~\ref{alg:determine})}{
     Roll back $S = S +\{i\}$\;
   }
 }
 \caption{Determine the state indices ($A7$) for the algorithm template.}
 \label{alg:state}
\end{algorithm}

The state initialization is generated based on the set of state indices using a heuristic. Consult Fig.~\ref{fig:sparse-dot-state}. The state is initialized directly above the first loop that is not part of the state indices. From this loop, any inner loops that are part of the state indices will have their value (tuple) associated with a state. Arrays (map) or scalars are chosen whether such inner loop indices exist, and the states are initialized accordingly.

\end{document}